\documentclass [ 10 pt ] {article}
\usepackage{tikz, pgf, pgfplots}
\usetikzlibrary
	{
		3d,
		arrows,											
		arrows.meta,
		automata,										
		backgrounds,
		bending,
		calc,
		chains,
		datavisualization.formats.functions,			
		decorations.pathmorphing,
		decorations.pathreplacing,
		decorations.text,
		fadings,
		fit,
		graphs,
		graphs.standard,
		matrix,
		positioning,									
		quantikz2,
		quotes,
		shadings,
		shadows.blur,
		shapes,
		trees
	}
\pgfplotsset{compat = newest}
\usepgflibrary
	{
		shapes.misc
	}
\usepgfplotslibrary
	{
		dateplot
	}
\usepackage{tikzpeople}
\usepackage{fontawesome5}

\usepackage{amsthm}
\usepackage{amssymb}
\usepackage[bb = boondox]{mathalfa}
\usepackage{dsfont}
\usepackage{newtxmath}
\usepackage{mathtools}
\usepackage{multicol}
\usepackage{braket}
\usepackage{physics}
\usepackage{empheq}

\usepackage[ compat = newest ]{yquant}

\usepackage{xcolor}
\usepackage{varwidth}
\usepackage[ skins, breakable, listings, theorems ] {tcolorbox}
\usepackage{graphicx}
\usepackage{xurl}                                                                                  %
\usepackage{hyperref}
\hypersetup
	{
		colorlinks,
		linkcolor = {WordRed!75!black},
		citecolor = {WordBlueDarker25},
		urlcolor = {WordAquaDarker50}
	}
\usepackage{colortbl}
\usepackage{float}
\usepackage{nicematrix}
\usepackage{cellspace}
\usepackage{makecell}
\usepackage{multirow}
\usepackage[ linesnumbered, tworuled, vlined ] {algorithm2e}
\usepackage{appendix}
\usepackage{enumitem}
\usepackage{siunitx}
\usepackage{xintexpr}
\usepackage{spreadtab}
\usepackage{framed}
\usepackage{svg}
\usepackage{lipsum}
\newcounter{mathseed}
\pgfmathsetseed{\arabic{mathseed}}
\pgfdeclarelayer{background}
\pgfsetlayers{background, main}
\pgfdeclaredecoration{irregular fractal line}{init}
{
	\state{init}[width=\pgfdecoratedinputsegmentremainingdistance]
	{
		\pgfpathlineto{\pgfpoint{random*\pgfdecoratedinputsegmentremainingdistance}{(random*\pgfdecorationsegmentamplitude-0.02)*\pgfdecoratedinputsegmentremainingdistance}}
		\pgfpathlineto{\pgfpoint{\pgfdecoratedinputsegmentremainingdistance}{0pt}}
	}
}
\def\tornpaper#1{%
	\ifthenelse{\isodd{\value{mathseed}}}
	{%
		\tikz
		{
			\node[inner sep = 1em] (A) {#1};		
			\begin{pgfonlayer}{background}			
				\fill[paper]						
				\pgfextra{\pgfmathsetseed{\arabic{mathseed}}\addtocounter{mathseed}{1}}%
				{decorate[irregular cloudy border]{decorate{decorate{decorate{decorate[ragged border]{
										(A.north west) -- (A.north east)
				}}}}}}
				-- (A.south east)
				\pgfextra{\pgfmathsetseed{\arabic{mathseed}}}%
				{decorate[irregular spiky border]{decorate{decorate{decorate{decorate[ragged border]{
										-- (A.south west)
				}}}}}}
				-- (A.north west);
			\end{pgfonlayer}
		}
	}
	{%
		\tikz{
			\node[inner sep=1em] (A) {#1};  
			\begin{pgfonlayer}{background}  
				\fill[paper] 
				\pgfextra{\pgfmathsetseed{\arabic{mathseed}}\addtocounter{mathseed}{1}}%
				{decorate[irregular spiky border]{decorate{decorate{decorate{decorate[ragged border]{
										(A.north east) -- (A.north west)
				}}}}}}
				-- (A.south west)
				\pgfextra{\pgfmathsetseed{\arabic{mathseed}}}%
				{decorate[irregular cloudy border]{decorate{decorate{decorate{decorate[ragged border]{
										-- (A.south east)
				}}}}}}
				-- (A.north east);
		\end{pgfonlayer}}
	}
}
\definecolor{MyLightRed}{RGB}{244, 213, 245}
\definecolor{WordRed}{RGB}{255, 0, 102}
\definecolor{WordRedAccent5Lighter60}{HTML}{F5B5A7}
\definecolor{WordRedAccent5Darker25}{HTML}{B23214}
\definecolor{RedDarkLightest}{HTML}{ff0088}
\definecolor{RedDarkLight}{HTML}{ea005f}
\definecolor{RedPurple}{HTML}{aa007f}
\definecolor{Purple}{HTML}{911146}
\definecolor{PurpleDark}{RGB}{102, 0, 102}
\definecolor{WordLightGreen}{RGB}{140, 214, 192}
\definecolor{WordGreen}{RGB}{0, 176, 80}
\definecolor{GreenLightest}{HTML}{00ffa0}
\definecolor{GreenLighter1}{HTML}{00b383}
\definecolor{GreenLighter2}{HTML}{00aa7f}
\definecolor{GreenDark}{HTML}{225522}
\definecolor{GreenTeal}{HTML}{008080}
\definecolor{WordIceBlue}{RGB}{223, 227, 229}
\definecolor{MyVeryLightBlue}{RGB}{211, 245, 247}
\definecolor{WordBlueVeryLight}{RGB}{0, 176, 240}
\definecolor{WordBlueLight}{RGB}{0, 112, 192}
\definecolor{WordBlueDark}{RGB}{46, 116, 181}
\definecolor{WordBlueDarker}{RGB}{31, 78, 121}
\definecolor{WordBlueDarker25}{RGB}{54, 96, 146}
\definecolor{WordBlueDarker50}{RGB}{36, 64, 98}
\definecolor{WordBlueDarkest}{RGB}{0, 32, 96}
\definecolor{WordBlue}{RGB}{19, 65, 99}
\definecolor{MyBlue}{RGB}{0, 64, 128}
\definecolor{MyDarkBlue}{RGB}{0, 51, 102}
\definecolor{BlueVeryDark}{HTML}{222255}
\definecolor{MagentaVeryLight}{RGB}{178, 162, 201}
\definecolor{MagentaLighter}{RGB}{161, 106, 221}
\definecolor{MagentaLight}{RGB}{128, 100, 162}
\definecolor{MagentaDark}{RGB}{106, 65, 152}
\definecolor{MagentaVeryDark}{RGB}{97, 75, 128}
\definecolor{WordAquaLighter80}{RGB}{218, 238, 243}
\definecolor{WordAquaLighter60}{RGB}{183, 222, 232}
\definecolor{WordAquaLighter40}{RGB}{146, 205, 220}
\definecolor{WordAquaDarker25}{RGB}{49, 134, 155}
\definecolor{WordAquaAccent2Darker25}{HTML}{398E98}
\definecolor{WordAquaDarker50}{RGB}{33, 89, 103}
\definecolor{WordVeryLightTeal}{RGB}{223, 236, 235}
\definecolor{WordLightTeal}{RGB}{160, 199, 197}
\definecolor{WordDarkTealLighter80}{RGB}{207, 223, 234}
\definecolor{WordDarkTeal}{RGB}{72, 123, 119}
\definecolor{WordDarkerTeal}{RGB}{48, 82, 80}
\definecolor{WordTurquoiseLighter80}{RGB}{209, 238, 249}
\definecolor{WordGoldAccent1Lighter40}{HTML}{FFDF6A}
\definecolor{WordGoldAccent1Darker25}{HTML}{C49A00}
\definecolor{Brown}{HTML}{666633}
\definecolor{WordOrangeAccent2Lighter60}{HTML}{FCD3A4}
\definecolor{WordOrangeAccent4Lighter60}{HTML}{F7C5A1}

\newtheorem{definition}{Definition}[section]

\usepackage
	[
		a4paper,
		left = 2.5cm,
		right = 2.5 cm,
		top = 2.5cm,
		bottom = 3.0 cm
	]
	{geometry}
\title
	{
		A Multiparty Quantum Private Equality Comparison scheme relying on $\ket{ GHZ_{ 3 } }$ states
	}

\usepackage{orcidlink}
\newcommand{\orcidicon}[1]{\href{https://orcid.org/#1}{\includegraphics[height=\fontcharht\font`\B]{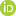}}}

\author
{
	Theodore Andronikos$^1$\orcidicon{0000-0002-3741-1271}
	and
	Alla Sirokofskich$^2$\\
	\\
	$^1$ \ Department of Informatics, Ionian University, \\
	7 Tsirigoti Square, 49100 Corfu, Greece; \\
	andronikos@ionio.gr \\
	$^2$ \ Department of History and Philosophy of Sciences, \\
	National and Kapodistrian University of Athens, \\
	Athens 15771, Greece; \\
	asirokof@math.uoa.gr
}
\begin{document}

\maketitle

\begin{abstract}
	This paper introduces an innovative entanglement-based protocol that accomplishes multiparty quantum private comparison leveraging maximally entangled $\ket{ GHZ_{ 3 } }$ triplets. The primary motivation behind this protocol is to design a protocol that can be readily executed by contemporary quantum computers. This is made possible because the protocol uses only $\ket{ GHZ_{ 3 } }$ triplets, irrespective of the number $n$ of millionaires. While more complex multi-particle entangled states, not to mention high dimensional quantum states, are possible, they are challenging to produce with existing quantum apparatus, leading to extended preparation time and complexity, particularly in scenarios involving numerous participants. By relying exclusively on $\ket{ GHZ_{ 3 } }$ states, which are the easiest to produce after Bell states, we avoid all these drawbacks, and take a decisive step towards the practical implementation of the GHZ$_{ 3 }$MQPEC protocol. An important quantitative characteristic of this protocol is that the required quantum resources are linear both in the number of millionaires and the volume of information to be compared. This further attests to the potential of the GHZ$_{ 3 }$MQPEC protocol for practical implementation. A notable aspect of the GHZ$_{ 3 }$MQPEC protocol is its suitability for both parallel and sequential execution. Although the execution of the quantum part of the protocol is envisioned to take place completely in parallel, it is also possible to be implemented sequentially. So, if the quantum resources do not suffice for the execution of the protocol in one go, it is possible to partition the millionaires into smaller groups and process these groups sequentially. Notably, our protocol involves two third parties; Trent, commonly featured in previous multiparty protocols, is now accompanied by Sophia. This dual setup allows simultaneous processing of all $n$ millionaires’ fortunes. Importantly, the GHZ$_{ 3 }$MQPEC protocol doesn’t rely on a quantum signature scheme or pre-shared keys, reducing complexity and cost. Implementation-wise, uniformity is ensured as all millionaires use similar private quantum circuits composed of Hadamard and CNOT gates, making it straightforward to execute on contemporary quantum computers. Lastly, the protocol is information-theoretically secure, preventing outside parties from learning about fortunes or inside players from knowing each other’s secret numbers.
	\\
\textbf{Keywords:}: Quantum private equality comparison, multiparty quantum private comparison,  $\ket{ GHZ_{ 3 } }$ states, semi-honest third party, quantum entanglement, quantum cryptography, quantum games.
\end{abstract}
\section{Introduction} \label{sec: Introduction}

The pursuit of quantum computers that surpass their classical counterparts remains ongoing. While we haven’t yet arrived at the promised land, recent milestones—such as IBM’s 127-qubit Eagle \cite{IBMEagle}, 433-qubit Osprey \cite{IBMOsprey}, and the colossal 1,121-qubit Condor \cite{IBMCondor}—indicate accelerated progress. Perhaps we are closer than anticipated to the quantum revolution. Given this context, it’s clear that quantum technology has matured significantly to warrant serious consideration for integration into a comprehensive framework designed to implement cryptographic protocols. In this perspective, it isn't surprising that quantum cryptographic methods have gained immense popularity since the landmark work by Bennett and Brassard \cite{Bennett1984}, where the first quantum key distribution protocol was presented. Today one can easily identify many active areas of research, such as quantum key distribution (see \cite{Bennett1984, Ekert1991, Bennett1992, Bennett2014} and the more recent \cite{Ampatzis2021}), quantum secret sharing \cite{Hillery1999, Cleve1999, Karlsson1999, Ampatzis2022}, quantum teleportation \cite{Bennett1993, Bouwmeester1997}, quantum secure direction communication \cite{Deng2003, Deng2004, Wang2005}, to name just a few.

Another notable direction that aims to achieve the tenets of privacy and security through quantum mechanics, is Quantum Private Comparison (QPC from now on), initiated by Yang and Wen in their pioneering work \cite{Yang2009}. Yang and Wen proposed the first quantum protocol that enables Alice and Bob to find out whether their private numbers are equal or not with the help of a third party. This is achieved in a way that no party gains any additional information in case the numbers are different. The origin of this problem can be traced back to the fundamental works \cite{Yao1982, Yao1986} by Yao, wherein the millionaires’ problem was introduced. In the millionaires’ problem, two millionaires wish to know who is richer, but without divulging any information about the precise amount of their fortune to the other party. Following the idea of the Yao’s millionaires’ problem, Boudot \cite{Boudot2001} proposed a classical protocol for the socialist millionaires’ problem, in which two millionaires want to know whether they happen to be equally rich. Yang and Wen's protocol is based on EPR pairs for comparing information with respect to equality, and employs the help of a third party (TP). Almost at the same time, the authors in \cite{Chen2010} proposed an alternative efficient QPC protocol based on GHZ triplets, also utilizing the help of a third party. In \cite{Liu2011} a QPC protocol based on W triplets, and also using a third party, was presented. Later, another QPC protocol relying on Bell States and a third party appeared in \cite{Tseng2011}.

In \cite{Liu2011a, Jia2011a, Liu2012, Liu2012a, Ji2016}, the authors proposed different QPC protocols based on entangled states, such as $\chi$, GHZ states, highly entangled six-qubit genuine states, that also rely on a semi-honest third party. The authors in \cite{Chou2016} presented the first semi-quantum private comparison protocol for participants who do not have any quantum capability to compare their secrets under an almost-dishonest third-party. Similarly, in \cite{He2023}, an improved semi-quantum private comparison protocol was introduced, achieving a higher security level. There have also been different approaches, e.g., the authors in \cite{Chen2012} do not use entanglement, but still require a semi-honest third party, as the QPC protocol proposed in \cite{Zi2013} that makes use of random rotation angles. In \cite{Ye2017} the first QPC protocol via cavity quantum electrodynamics was proposed. Recently, in \cite{Hou2024} a practical quantum secure protocol for the socialist millionaires’ problem based on single photons was introduced, with emphasis on ease of implementation with current technology. For more pointers on the subject, the interested reader may consult the comprehensive survey \cite{Liu2013}. Sophisticated quantum protocols for magnitude comparison have also been proposed in \cite{Jia2011, Lin2012, Zhang2013, Ye2018, Cao2019, Wu2021}; all these protocols envision $d$-level quantum systems, and also require a third party. Typically, $d$ is assumed to be greater than a certain threshold, which, in practice, considerably complicates their implementation on contemporary quantum computers based on qubits.

Chang et al. in \cite{Chang2012} were the first to devise a clever generalization to $n$ parties that requires $\ket{ GHZ_{ n } }$-class states and the prior distribution of a secret key to each of the $n$ parties. This inspired researchers to design more multiparty quantum private comparison (MQPC for short) protocols. Along this direction, the authors in \cite{Liu2013a} proposed a different MQPC protocol based on a $d$-level quantum system. Similarly, the researchers in \cite{Huang2015} introduced a MPQC protocol with an almost-dishonest third party that requires a $d$-level quantum system, where also $d$ is assumed greater than the largest number slated for comparison. Subsequently, in \cite{Hung2016} an enhanced MQPC protocol was introduced that uses $\ket{ GHZ_{ n } }$-type states for $n$ participants. Ye and Ye in \cite{Ye2018} came up with a sophisticated MQPC protocol also based on a $d$-level quantum system. One of the novelties of their protocol was the creative utilization of two semi-honest third parties that eliminate the need of pre-shared keys between each pair. Other notable MQPC protocols were proposed in \cite{Cao2019}, relying on $d$-level $\ket{ GHZ_{ n } }$ states, where $n$ is the number of parties, and in \cite{Zhang2023} exploiting properties of quantum homomorphic encryption.

In constructing secure QPC protocols, there are certain obstacles. Colbeck showed that unconditionally secure two-party classical computation is impossible for many classes of functions \cite{Colbeck2007}. Despite the fact that things are different in the quantum setting \cite{Crepeau2002}, Lo has pointed out that the equality function cannot be securely evaluated with a two-party scenario \cite{Lo1997}. Therefore, some additional assumptions (e.g., semi-honest parties) should be considered in order to attain the goal of secure private comparison.

The proposed protocol is described as a quantum game, something quite usual in the literature. The role of the millionaires is undertaken by the famous Alice and her clones, which star alongside Sophia and Trent that play the two third parties that are necessary for the implementation of the protocol. Employing games in our presentation aims to facilitate comprehension of technical concepts. Quantum games, since their inception in 1999 \cite{Meyer1999, Eisert1999}, have gained widespread acceptance. This is not surprising, considering the potential superiority of quantum strategies over classical ones \cite{Andronikos2018, Andronikos2021, Andronikos2022a, Kastampolidou2023a}. Notably, the renowned prisoners' dilemma game exemplifies this phenomenon, extending to other abstract quantum games as well \cite{Eisert1999, Giannakis2019}. Moreover, the quantization of various classical systems can be applied to political structures, as demonstrated recently in \cite{Andronikos2022}. In the realm of quantum cryptography, the presentation of protocols often takes the form of games, a common practice evident in recent works such as \cite{Ampatzis2021, Ampatzis2022, Ampatzis2023, Andronikos2023, Andronikos2023a, Andronikos2023b, Karananou2024, Andronikos2024}. Perhaps unexpectedly, the quantization of classical systems finds applications even in political structures \cite{Andronikos2022}. In the broader context of game-theoretic applications, unconventional environments, such as biological systems, have garnered significant attention \cite{Theocharopoulou2019, Kastampolidou2020a, Kostadimas2021}. It's intriguing to note that biological systems may give rise to biostrategies that outperform classical ones, even in iconic games like the Prisoners' Dilemma \cite{Kastampolidou2020, Kastampolidou2021, Kastampolidou2023, Papalitsas2021, Adam2023}.

\textbf{Contribution}. This paper introduces an innovative entanglement-based protocol, aptly named GHZ$_{ 3 }$MQPEC, that accomplishes multiparty quantum private comparison leveraging maximally entangled $\ket{ GHZ_{ 3 } }$ triplets. The primary motivation behind this protocol is to design a protocol that can be readily executed by contemporary quantum computers. This is made possible because the protocol uses only $\ket{ GHZ_{ 3 } }$ triplets, irrespective of the number $n$ of millionaires. While more complex multi-particle entangled states, not to mention high dimensional quantum states, are possible, they are challenging to produce with existing quantum apparatus, leading to extended preparation time and complexity, particularly in scenarios involving numerous participants. Contemporary quantum computers can generate $\ket{ GHZ_{ n } }$ states for small values of $n$. Unfortunately, the preparation and distribution of these states become increasingly difficult as $n$ grows. By relying exclusively on $\ket{ GHZ_{ 3 } }$ states, which are the easiest to produce after Bell states, we avoid all these drawbacks, and take a decisive step towards the practical implementation of the GHZ$_{ 3 }$MQPEC protocol.

An important quantitative characteristic of this protocol is that, as shown in subsection \ref{subsec: Efficiency}, the required quantum resources are linear both in the number of millionaires $n$ and in the number $m$ of qubits within each quantum register, which reflects the magnitude of the numbers to be compared. This allows its seamless scalability not only in participant count but also in the amount of information processed. This further attests to the potential of the GHZ$_{ 3 }$MQPEC protocol for practical implementation.

A notable aspect of the GHZ$_{ 3 }$MQPEC protocol is its suitability for both parallel and sequential execution. Although the execution of the quantum part of the protocol is envisioned to take place completely in parallel, it is also possible to be implemented sequentially. So, if the quantum resources do not suffice for the execution of the protocol in one go, it is possible to partition the $n$ of millionaires into smaller groups and process these groups sequentially, even to the point of executing the protocol serially for one millionaire at a time. Another distinguishing feature of our protocol is the existence of two third parties, so Trent, featured in many previous multiparty protocols, is now accompanied by Sophia. The existence of two parties enables the simultaneous processing of all $n$ millionaires' fortunes in one go. In addition to the possibility of either parallel or sequential execution, this protocol is flexible enough to accommodate both localized and distributed settings, a capability that stems from its inherent use of entanglement. The localized scenario involves all players physically residing at the same spatial location simultaneously. In contrast, the distributed scenario accommodates players located at different places.

It is worth mentioning that the GHZ$_{ 3 }$MQPEC protocol doesn't rely on a quantum signature scheme, or pre-shared keys, which reduces its complexity and cost. Furthermore, in terms of its implementation, the protocol ensures uniformity as all millionaires employ identical private quantum circuits. This design establishes a completely modular quantum system where all modules are identical. Each private quantum circuit exclusively employs the widely-used Hadamard and CNOT quantum gates, facilitating straightforward implementation on contemporary quantum computers. Finally, it goes without saying, that outside parties cannot learn any information about the millionaires’ fortunes, inside players cannot know the secret numbers of other players, and the GHZ$_{ 3 }$MQPEC protocol is shown to be information-theoretically secure.

\subsection*{Organization} \label{subsec: Organization}

This article is organized as follows. The Introduction (section \ref{sec: Introduction}) presents the subject matter, accompanied by bibliographic references to related works. A concise overview of the essential concepts is provided in section \ref{sec: Preliminary Concepts}, laying the foundation for understanding our protocol. A detailed explanation of the hypotheses underlying the  GHZ$_{ 3 }$MQPEC protocol is given in section \ref{sec: The Setup}. The GHZ$_{ 3 }$MQPEC protocol is formally presented in section \ref{sec: The GHZ$_{ 3 }$MQPEC Protocol}. A detailed example illustrating the inner workings of the protocol is given in section \ref{sec: The GHZ$_{ 3 }$MQPEC Protocol in Action}, while efficiency and security issues are analyzed in section \ref{sec: Efficiency & Security Analysis}. The paper concludes with a summary and discussion of the finer points of the protocol in section \ref{sec: Discussion and Conclusions}.

\section{Preliminary concepts} \label{sec: Preliminary Concepts}

\subsection{$\ket{ GHZ_{ 3 } }$ states} \label{subsec: ket{ GHZ_{ 3 } } States}

Many quantum protocols can be described as games between well-known fictional players, commonly referred to as Alice, Bob, Charlie, etc. and in this particular case Sophia and Trent besides Alice. Our protagonists, while being spatially separated, are attempting to exchange information privately and securely. Typically, secure communication can be established via a combination of classical pairwise authenticated channels with pairwise quantum channels. Usually, the process of transmitting secret information takes place through the quantum channel using a multitude of different techniques, and, subsequently, in order to complete the process, messages are exchanged through a classical public channel. During this phase, an adversary who is mostly referred to as Eve, may appear and attempt to track this communication and steal any information possible. In such an eventuality, the major advantage of quantum cryptography over its classical counterpart is that during the transmission of information through the quantum channel, the communicating players are protected due to certain fundamental principles of quantum mechanics, such as the no-cloning theorem \cite{wootters1982single}, entanglement monogamy, to name the most prominent.

Entanglement constitutes the fundamental basis of most quantum protocols, including the one proposed in this work, and possibly the de facto future of quantum cryptography, due to its numerous applications in the entire field. It is one of the fundamental principles of quantum mechanics and can be described mathematically as the linear combination of two or more product states. The entanglement of three or more qubits is referred to as a GHZ state. The fundamental idea of quantum entanglement in its simplest form is that it is possible for quantum particles to be entangled together and when a property is measured in one particle, it can be observed on the other particles instantaneously. The reader who wants to dive deeper into this phenomenon may consult any standard textbook, such as \cite{Nielsen2010, Yanofsky2013a, Wong2022}, for a more detailed exposition.

Contemporary quantum computers are powerful enough (see for instance the recent IBM quantum computers \cite{IBMOsprey, IBMCondor}) to be able to generate GHZ states utilizing standard quantum gates like the Hadamard and CNOT gates. Implementing such a circuit requires $n$ qubits, one Hadamard gate that is applied to the first qubit, and $n - 1$ CNOT gates. We refer to \cite{Cruz2019} for a practical methodology that can be utilized to construct efficient GHZ circuits, in the sense that it just takes $\lg n$ steps to produce the $\ket{ GHZ_{ n } }$ state. The proposed protocol makes use of $\ket{ GHZ_{ 3 } }$ triplets. The mathematical description of the $\ket{ GHZ_{ 3 } }$ state is given below. To make clear that $\ket{ GHZ_{ 3 } }$ entanglement requires three separate entities, we use subscripts A, S, and T to indicate which qubit belongs to Alice, Sophie and Trent. Following this convention, qubits denoted by $\ket{ \cdot }_{ A }$ belong to Alice, those denoted by $\ket{ \cdot }_{ S }$ belong to Sophie, and those denoted by $\ket{ \cdot }_{ T }$ belong to Trent.
\begin{align} \label{eq: GHZ_{ 3 } State}
	\ket{ GHZ_{ 3 } }
	=
	\frac { 1 } { \sqrt{ 2 } }
	\left(
	\ket{ 0 }_{ T }
	\ket{ 0 }_{ S }
	\ket{ 0 }_{ A }
	+
	\ket{ 1 }_{ T }
	\ket{ 1 }_{ S }
	\ket{ 1 }_{ A }
	\right)
	\ .
\end{align}
Our protocol requires not just a single $\ket{ GHZ_{ 3 } }$ triplet, but $m$ such triplets. The state of a composite system comprised of $m$ $\ket{ GHZ_{ 3 } }$ triplets is given by the next formula (for details we refer to \cite{Ampatzis2022} and \cite{Ampatzis2023}).
\begin{align} \label{eq: m-Fold GHZ_{ 3 } Triplets}
	\ket{ GHZ_{ 3 } }^{ \otimes m }
	=
	\frac { 1 } { \sqrt{ 2^{ m } } }
	\sum_{ \mathbf{ x } \in \mathbb{ B }^{ m } }
	\ket{ \mathbf{ x } }_{ T }
	\ket{ \mathbf{ x } }_{ S }
	\ket{ \mathbf{ x } }_{ A }
	\ .
\end{align}
In writing formula \eqref{eq: m-Fold GHZ_{ 3 } Triplets} the following notation is used.

\begin{itemize}
	[ left = 0.00 cm, labelsep = 0.50 cm ]
	\item	$\mathbb{ B }$ is the binary set $\{ 0, 1 \}$.
	\item	To syntactically distinguish the bit vector $\mathbf{ x } \in \mathbb{ B }^{ m }$ from the bit $x \in \mathbb{ B }$, we write the bit vector $\mathbf{ x }$ in boldface. A bit vector $\mathbf{ x }$ of length $m$ is a sequence of $n$ bits: $\mathbf{ x } = x_{ m - 1 } \dots x_{ 0 }$. There is also the special zero bit vector $\mathbf{ 0 }$, in which all the bits are zero, i.e., $\mathbf{ 0 } = 0 \dots 0$.
	\item	Each bit vector $\mathbf{ x } \in \mathbb{ B }^{ m }$ serves as the binary representation of one of the $2^{ m }$ basis kets that form the computational basis of the underlying $2^{ m }$-dimensional Hilbert space.
	\item	As we have previously explained, to avoid ambiguity we rely on the subscripts A, S and T to indicate the qubits that belong to Alice, Sophie and Trent's quantum registers respectively.
\end{itemize}

\subsection{Inner product modulo $2$ operation} \label{subsec: Inner Product Modulo $2$ Operation}

\emph{Inner product modulo} $2$ is a very useful binary operation defined on bit vectors of equal length. The inner product modulo $2$ takes as inputs two bit vectors $\mathbf{ x }, \mathbf{ y } \in \mathbb{ B }^{ m }$, and outputs their inner product, denoted by $\mathbf{ x \bullet y }$. Assuming that $\mathbf{ x } = x_{ m - 1 } \dots x_{ 0 }$ and $\mathbf{ y } = y_{ m - 1 } \dots y_{ 0 }$, then $\mathbf{ x } \bullet \mathbf{ y }$ is given by
\begin{align} \label{eq: Inner Product Modulo $2$}
	\mathbf{ x }
	\bullet
	\mathbf{ y }
	\coloneq
	x_{ m - 1 } y_{ m - 1 }
	\oplus \dots \oplus
	x_{ 0 } y_{ 0 }
	\ ,
\end{align}
where the symbol $\coloneq$ stands for ``is defined as,'' and $\oplus$ is \emph{addition modulo} $2$. This operation is employed in many important formulae of quantum computation and information. One such fundamental formula is given below; it concerns the $m$-fold Hadamard transform of an arbitrary basis ket $\ket{ \mathbf{ x } }$ and we shall use it during the detailed analysis of our quantum protocol. The proof can be found in most standard textbooks, e.g., \cite{Mermin2007, Nielsen2010}.
\begin{align} \label{eq: Hadamard m-Fold Ket x}
	H^{ \otimes m } \ket{ \mathbf{ x } }
	&=
	\frac { 1 } { \sqrt{ 2^{ m } } }
	\sum_{ \mathbf{ z } \in \mathbb{ B }^{ m } }
	( - 1 )^{ \mathbf{ z \bullet x } } \ket{ \mathbf{ z } }
	\ .
\end{align}
A particularly useful property of the inner product modulo $2$ operation is a special type of symmetry that can be used to partition the set of bit vectors into two subsets of equal cardinality. Specifically, any fixed element $\mathbf{ c }$ of $\mathbb{ B }^{ m }$ different from $\mathbf{ 0 }$, exhibits the following property: for exactly half of the elements $\mathbf{ x } \in \mathbb{ B }^{ m }$, $\mathbf{c} \bullet \mathbf{ x } = 0$, and for the remaining half $\mathbf{ c } \bullet \mathbf{ x } = 1$. Clearly, $\mathbf{ 0 }$ is an exception to the above rule because if $\mathbf{ c } = \mathbf{ 0 }$, then for all $\mathbf{ x } \in \mathbb{ B }^{ m }$, $\mathbf{ c } \bullet \mathbf{ x } = 0$. Following \cite{Andronikos2023b}, we call this property the Characteristic Inner Product (CIP) property.
		\begin{align}
			\mathbf{ c } = \mathbf{ 0 }
			&\Rightarrow
			\text{for all } 2^{ m } \text{ bit vectors } \mathbf{ x } \in \mathbb{ B }^{ m },
			\text{ } \mathbf{ c } \bullet \mathbf{ x } = 0
			\label{eq: Inner Product Modulo $2$ Property For Zero}
			\\
			\mathbf{ c } \neq \mathbf{ 0 }
			&\Rightarrow
			\left\{
			\
			\begin{matrix*}[l]
				\text{for } 2^{ m - 1 } \text{ bit vectors } \mathbf{ x } \in \mathbb{ B }^{ m }, \ \mathbf{ c } \bullet \mathbf{ x } = 0
				\\
				\text{for } 2^{ m - 1 } \text{ bit vectors } \mathbf{ x } \in \mathbb{ B }^{ m }, \ \mathbf{ c } \bullet \mathbf{ x } = 1
			\end{matrix*}
			\
			\right\}
			\label{eq: Inner Product Modulo $2$ Property For NonZero}
		\end{align}

For our protocol, and also the purpose of creating decoys, we utilize two other signature states, namely $\ket{ + }$ and $\ket{ - }$, defined as
\begin{tcolorbox}
	[
	enhanced,
	breakable,
	grow to left by = 0.00 cm,
	grow to right by = 0.00 cm,
	colback = white,			
	enhanced jigsaw,			
	sharp corners,
	toprule = 0.1 pt,
	bottomrule = 0.1 pt,
	leftrule = 0.1 pt,
	rightrule = 0.1 pt,
	sharp corners,
	center title,
	fonttitle = \bfseries
	]
	\begin{minipage}[b]{0.45 \textwidth}
		\begin{align} \label{eq: Ket +}
			\ket{ + } = H \ket{ 0 } = \frac { \ket{ 0 } + \ket{ 1 } } { \sqrt{ 2 } }
		\end{align}
	\end{minipage} 
	\hfill
	\begin{minipage}[b]{0.45 \textwidth}
		\begin{align} \label{eq: Ket -}
			\ket{ - } = H \ket{ 1 } = \frac { \ket{ 0 } - \ket{ 1 } } { \sqrt{ 2 } }
		\end{align}
	\end{minipage}
\end{tcolorbox}
Although during the execution of the protocol itself, measurements are performed with respect to the computational basis $\{ \ket{ 0 }, \ket{ 1 } \}$, during the test for eavesdropping detection, decoys are also measured with respect to the Hadamard basis.

\section{The setup} \label{sec: The Setup}

This paper introduces a novel protocol designed to practically and efficiently accomplish multiparty quantum private equality comparison using only $\ket{ GHZ_{ 3 } }$ states. Henceforth, for succinctness, we shall refer to this protocol with the acronym GHZ$_{ 3 }$MQPEC. The current section is devoted to the precise explanation of the whole setup and the hypotheses that guarantee the correct implementation of GHZ$_{ 3 }$MQPEC.

\subsection{The actors} \label{subsec: The Actors}

Before we formally introduce the players involved in the GHZ$_{ 3 }$MQPEC, we clarify one important notion, that of the \emph{semi-honest} player. After it was established that the equality function cannot be securely evaluated with a two-party scenario \cite{Lo1997}, it became clear that additional assumptions must be introduced in order to achieve secure private comparison. In the literature, the most well-known such assumption is that of a semi-honest third party. Definition \ref{def: Semi-Honest Player} clarifies what this entails.

\begin{definition} [Semi-honest player] \label{def: Semi-Honest Player}
	A \emph{semi-honest} player exhibits the following traits:
	\begin{itemize}
		[ left = 0.00 cm, labelsep = 0.50 cm ]
		\item	Always executes the protocol faithfully.
		\item	Cannot conspire with any other player.
		\item	Cannot be corrupted by an outside eavesdropper.
		\item	Records all intermediate computations and might try to steal information from the records.
	\end{itemize}
\end{definition}

In other words, a semi-honest player sticks to the rules of the protocol so as to help the interested parties do the equality comparison, but, at the same time, is also curious to find out their private numbers.

The GHZ$_{ 3 }$MQPEC protocol evolves as a game played by $n + 2$ players, where $n$ is an arbitrarily large positive integer. So, without further ado, we list the actors and the rules governing their behavior below.

\begin{enumerate}
	\renewcommand\labelenumi{(\textbf{A}$_{ \theenumi }$)}
	\item	There are $n$ players, the millionaires, who are designated by Alice$_{ 0 }$, \dots, Alice$_{ n - 1 }$. Each Alice$_{ i }$, $0 \leq i \leq n - 1$, desires to find out whether her fortune is equal or not to the fortune of every other Alice$_{ j }$, $0 \leq j \neq i \leq n - 1$, but \textit{without ever divulging her actual fortune} to any other in-game player or outside eavesdropper. For the needs of the protocol, the millionaires' fortunes are represented in binary and not in decimal. In particular, Alice$_{ i }$'s fortune is designated by the bit vector $\mathbf{ f }_{ i }$, $0 \leq i \leq n - 1$. There is a catch, however, in the whole setup. The millionaires, in case their fortunes turn out to be different, must not be able to infer any additional information besides the fact that they are different, not even a single bit of the other player’s fortune.
	
	To allow for more generality the millionaires are assumed to be in different locations.
	\item	Additionally, there are $2$ semi-honest third parties, Sophia and Trent, whose participation is essential for the correct implementation of the protocol. So, in total, this game involves $n + 2$ players, all in different regions of space.
	\item	Sophia's role is critical as we explain now. The rationale behind the GHZ$_{ 3 }$MQPEC protocol is the following: instead of Alice$_{ i }$ and Alice$_{ j }$ directly comparing their fortunes $\mathbf{ f }_{ i }$ and $\mathbf{ f }_{ j }$, $0 \leq j \neq i \leq n - 1$, they both compare their fortune with a fixed, but secret number $\mathbf{ s }$. If $\mathbf{ s } \oplus \mathbf{ f }_{ i } = \mathbf{ s } \oplus \mathbf{ f }_{ j }$, then $\mathbf{ f }_{ i } = \mathbf{ f }_{ j }$, whereas if $\mathbf{ s } \oplus \mathbf{ f }_{ i } \neq \mathbf{ s } \oplus \mathbf{ f }_{ j }$, then $\mathbf{ f }_{ i } \neq \mathbf{ f }_{ j }$. Sophia is responsible for randomly choosing a secret number $\mathbf{ s }$ and embedding it into the global entanglement. It goes without saying that Sophia \textit{must never reveal her secret number} $\mathbf{ s }$ to any other in-game player or outside eavesdropper.
	\item	Trent's role is also essential. First of all, he is responsible for the generation of the $\ket{ GHZ_{ 3 } }$ triplets and their distribution to the $n + 2$ players according to the distribution scheme outlined in Definition \ref{def: Entanglement Distribution Scheme}. Moreover, at the end of the protocol, he determines for each Alice$_{ i }$, $0 \leq i \leq n - 1$, whether her fortune is equal or not to that of every other Alice$_{ j }$, $0 \leq j \neq i \leq n - 1$. Trent must only announce the corresponding decision in the form of a single bit, e.g., YES$_{ i, j }$ or NO$_{ i, j }$. \textit{Trent must never reveal the private measurements and records} he used to arrive at his decisions to any other in-game player or outside eavesdropper.
	\item	All players agree beforehand on the number $m$ of qubits sufficient to store their fortunes $\mathbf{ f }_{ i }$, $0 \leq i \leq n - 1$, and Sophia's secret number $\mathbf{ s }$.
\end{enumerate}

We clarify that the millionaires do not wish to compare their fortunes in terms of relative size, i.e., to establish which one is greater than the other; just to know if they are equal or not. Furthermore, as is the norm in the literature, we assume that the quantum channel is an ideal channel, meaning that there is no noise and the particle is not lost, that the classical channel is an authenticated channel where the transmitted message is public but cannot be modified by an adversary.

\subsection{The entanglement distribution scheme} \label{subsec: The Entanglement Distribution Scheme}

The implementation of the GHZ$_{ 3 }$MQPEC protocol requires a versatile composite system consisting of smaller subsystems. The distinguishing characteristic of this setting is that the corresponding qubits across all triplets of associated quantum registers are entangled in the $\ket{ GHZ_{ 3 } }$ state. This is an immediate consequence of the employed Uniform $\ket{ GHZ_{ 3 } }$ Distribution Scheme, as explained in the next Definition \ref{def: Entanglement Distribution Scheme}.

\begin{definition} [Uniform $\ket{ GHZ_{ 3 } }$ Distribution Scheme] \label{def: Entanglement Distribution Scheme}
	The Uniform $\ket{ GHZ_{ 3 } }$ Distribution Scheme stipulates the following.
	\begin{itemize}
		[ left = 0.00 cm, labelsep = 0.50 cm ]
		\item	
		Alice$_{ i }$, $0 \leq i \leq n - 1$, is endowed with an $m$-qubit input register $AIR_{ i }$.
		\item	
		For every Alice$_{ i }$, Sophia and Trent each utilize an $m$-qubit input register corresponding to Alice$_{ i }$, denoted by $SIR_{ i }$ and $TIR_{ i }$, $0 \leq i \leq n - 1$, respectively.
		\item	
		For every Alice$_{ i }$, $0 \leq i \leq n - 1$, Trent generates $3m$ triplets $( p, q, r )$ entangled in the $\ket{ GHZ_{ 3 } }$ state. Trent keeps the first qubit $p$, and transmits the second and third qubits $q$ and $r$ to Sophia and Alice$_{ i }$, respectively.
	\end{itemize}
\end{definition}

This scheme effectively creates for every Alice$_{ i }$, $0 \leq i \leq n - 1$, a triplet of quantum registers, namely $(TIR_{ i }, SIR_{ i }, AIR_{ i })$. These associated registers are strongly linked because their corresponding qubits are maximally entangled in the $\ket{ GHZ_{ 3 } }$ state. This is visualized in Figure \ref{fig: The Entanglement Distribution Scheme} that depicts this setup, with the corresponding qubits that form the $\ket{ GHZ_{ 3 } }$ triplet colored in the same color. In this composite system each register holds $m$ qubits. We point out that it doesn't matter in the least if the registers are all in the same place, or are all in different spatial locations. The power of the entanglement effect, stemming from the $m$ $\ket{ GHZ_{ 3 } }$ triplets, will instill the necessary correlation, irrespective of whether the composite system is localized or entirely distributed. It is precisely this unique effect of entanglement that allows us to envision the whole setting as a unified system.

Any practical implementation of the GHZ$_{ 3 }$MQPEC protocol will require Trent to also prepare decoys. This technique is ubiquitous in the literature and is, typically, referred to as the \emph{decoy technique}. There are many well-written exposition that thoroughly explain the specifics; we refer to \cite{Deng2008, Yang2009, Tseng2011, Chang2012, Hung2016, Ye2018, Wu2021, Hou2024}, to cite just a few. The idea is that, in order to detect the presence of a possible eavesdropper, the party responsible for the generation of the entangled triplets also produces decoys, randomly chosen from one of the $\ket{ 0 }, \ket{ 1 }, \ket{ + }, \ket{ - }$ states. These decoys are randomly inserted into the transmitted sequence(s). We have not mentioned the decoys in the Definition \ref{def: Entanglement Distribution Scheme} to avoid clutter, and make the main idea of the protocol easier to grasp.

\begin{tcolorbox}
	[
	enhanced,
	breakable,
	grow to left by = 0.00 cm,
	grow to right by = 0.00 cm,
	colback = MagentaLight!12,			
	enhanced jigsaw,					
	sharp corners,
	toprule = 1.0 pt,
	bottomrule = 1.0 pt,
	leftrule = 0.1 pt,
	rightrule = 0.1 pt,
	sharp corners,
	center title,
	fonttitle = \bfseries
	]
	\begin{figure}[H]
		\centering
		\begin{tikzpicture} [ scale = 0.750, transform shape ]
			\node
			[
			maninblack,
			scale = 1.50,
			anchor = center,
			label = { [ label distance = 0.00 cm ] below: Trent }
			]
			(Trent) { };
			\matrix
			[
			matrix of nodes,
			nodes in empty cells,
			column sep = 1.00 mm,
			right = 0.750 of Trent,
			nodes = { circle, minimum size = 10 mm, semithick, font = \footnotesize },
			]
			{
				\node [ shade, outer color = RedPurple!75, inner color = white ] (T_n-1_m-1) { $m - 1$ }; &
				\node [ shade, outer color = WordAquaLighter60, inner color = white ] (T_n-1_Dots) { \large \dots }; &
				\node [ shade, outer color = GreenLighter2!75, inner color = white ] (T_n-1_0) { $0$ };
				\\
			};
			\node
			[
			right = 5.00 of Trent
			]
			(Trent_Dots) { \LARGE \dots };
			\matrix
			[
			matrix of nodes,
			nodes in empty cells,
			column sep = 1.00 mm,
			right = 6.50 of Trent,
			nodes = { circle, minimum size = 10 mm, semithick, font = \footnotesize },
			]
			{
				\node [ shade, outer color = RedPurple!75, inner color = white ] (T_0_m-1) { $m - 1$ }; &
				\node [ shade, outer color = WordAquaLighter60, inner color = white ] (T_0_Dots) { \large \dots }; &
				\node [ shade, outer color = GreenLighter2!75, inner color = white ] (T_0_0) { $0$ };
				\\
			};
			\node
			[
			alice,
			scale = 1.50,
			anchor = center,
			above = 3.00 cm of Trent,
			label = { [ label distance = 0.00 cm ] below: Sophia }
			]
			(Sophia) { };
			\matrix
			[
			matrix of nodes,
			nodes in empty cells,
			column sep = 1.00 mm,
			right = 0.750 of Sophia,
			nodes = { circle, minimum size = 10 mm, semithick, font = \footnotesize },
			]
			{
				\node [ shade, outer color = RedPurple!75, inner color = white ] (S_n-1_m-1) { $m - 1$ }; &
				\node [ shade, outer color = WordAquaLighter60, inner color = white ] (S_n-1_Dots) { \large \dots }; &
				\node [ shade, outer color = GreenLighter2!75, inner color = white ] (S_n-1_0) { $0$ };
				\\
			};
			\node
			[
			right = 5.00 of Sophia
			]
			(Sophia_Dots) { \LARGE \dots };
			\matrix
			[
			matrix of nodes,
			nodes in empty cells,
			column sep = 1.00 mm,
			right = 6.50 of Sophia,
			nodes = { circle, minimum size = 10 mm, semithick, font = \footnotesize },
			]
			{
				\node [ shade, outer color = RedPurple!75, inner color = white ] (S_0_m-1) { $m - 1$ }; &
				\node [ shade, outer color = WordAquaLighter60, inner color = white ] (S_0_Dots) { \large \dots }; &
				\node [ shade, outer color = GreenLighter2!75, inner color = white ] (S_0_0) { $0$ };
				\\
			};
			\node
			[
			charlie,
			female,
			scale = 1.50,
			anchor = center,
			above = 3.00 cm of Sophia,
			label = { [ label distance = 0.00 cm ] below: Alice$_{ n - 1 }$ }
			]
			(Alice_n-1) { };
			\matrix
			[
			matrix of nodes,
			nodes in empty cells,
			column sep = 1.00 mm,
			right = 0.750 of Alice_n-1,
			nodes = { circle, minimum size = 10 mm, semithick, font = \footnotesize },
			]
			{
				\node [ shade, outer color = RedPurple!75, inner color = white ] (A_n-1_m-1) { $m - 1$ }; &
				\node [ shade, outer color = WordAquaLighter60, inner color = white ] (A_n-1_Dots) { \large \dots }; &
				\node [ shade, outer color = GreenLighter2!75, inner color = white ] (A_n-1_0) { $0$ };
				\\
			};
			\node
			[
			right = 5.00 of Alice_n-1
			]
			(Alice_Dots) { \LARGE \dots };
			\matrix
			[
			matrix of nodes,
			nodes in empty cells,
			column sep = 1.00 mm,
			right = 6.50 of Alice_n-1,
			nodes = { circle, minimum size = 10 mm, semithick, font = \footnotesize },
			]
			{
				\node [ shade, outer color = RedPurple!75, inner color = white ] (A_0_m-1) { $m - 1$ }; &
				\node [ shade, outer color = WordAquaLighter60, inner color = white ] (A_0_Dots) { \large \dots }; &
				\node [ shade, outer color = GreenLighter2!75, inner color = white ] (A_0_0) { $0$ };
				\\
			};
			\node
			[
			above right = 3.25 cm and 5.00 cm of Alice_n-1,
			anchor = center,
			shade,
			top color = GreenTeal, bottom color = black,
			rectangle,
			text width = 10.00 cm,
			align = center
			]
			(Label)
			{ \color{white} \textbf{The entangled triplets of qubits distributed by Trent to Sophia and the millionaires. The characteristic property of this system is that all qubits in the corresponding positions $0, 1, \dots, m - 1$ make up a $\ket{ GHZ_{ 3 } }$ triplet.} };
			\node
			[
			draw = WordBlueDarker50,
			line width = 1.5pt,
			dash pattern = on 1pt off 4pt on 6pt off 4pt, inner sep = 2mm,
			rectangle,
			rounded corners,
			fit = (T_n-1_m-1) (T_n-1_0),
			label = { [ label distance = 0.00 cm ] north: \color{WordBlueDarker50} $TIR_{ n - 1 }$ }
			]
			( ) { };
			\node
			[
			draw = WordBlueDarker50,
			line width = 1.5pt,
			dash pattern = on 1pt off 4pt on 6pt off 4pt, inner sep = 2mm,
			rectangle,
			rounded corners,
			fit = (T_0_m-1) (T_0_0),
			label = { [ label distance = 0.00 cm ] north: \color{WordBlueDarker50} $TIR_{ 0 }$ }
			]
			( ) { };
			\node
			[
			draw = WordBlueDarker50,
			line width = 1.5pt,
			dash pattern = on 1pt off 4pt on 6pt off 4pt, inner sep = 2mm,
			rectangle,
			rounded corners,
			fit = (S_n-1_m-1) (S_n-1_0),
			label = { [ label distance = 0.00 cm ] north: \color{WordBlueDarker50} $SIR_{ n - 1 }$ }
			]
			( ) { };
			\node
			[
			draw = WordBlueDarker50,
			line width = 1.5pt,
			dash pattern = on 1pt off 4pt on 6pt off 4pt, inner sep = 2mm,
			rectangle,
			rounded corners,
			fit = (S_0_m-1) (S_0_0),
			label = { [ label distance = 0.00 cm ] north: \color{WordBlueDarker50} $SIR_{ 0 }$ }
			]
			( ) { };
			\node
			[
			draw = WordBlueDarker50,
			line width = 1.5pt,
			dash pattern = on 1pt off 4pt on 6pt off 4pt, inner sep = 2mm,
			rectangle,
			rounded corners,
			fit = (A_n-1_m-1) (A_n-1_0),
			label = { [ label distance = 0.00 cm ] north: \color{WordBlueDarker50} $AIR_{ n - 1 }$ }
			]
			( ) { };
			\node
			[
			draw = WordBlueDarker50,
			line width = 1.5pt,
			dash pattern = on 1pt off 4pt on 6pt off 4pt, inner sep = 2mm,
			rectangle,
			rounded corners,
			fit = (A_0_m-1) (A_0_0),
			label = { [ label distance = 0.00 cm ] north: \color{WordBlueDarker50} $AIR_{ 0 }$ }
			]
			( ) { };
			\begin{scope}[on background layer]
				\node
				[
				above = - 1.50 cm of T_n-1_m-1,
				rectangle,
				rounded corners = 8 pt,
				fill = RedPurple!25,
				minimum width = 6 mm,
				minimum height = 95 mm
				]
				( ) { };
				\node
				[
				below = 0.750 cm of T_n-1_m-1
				]
				( ) {\colorbox{RedPurple!75} {$\ket{ GHZ_{ 3 } }$}};
				\node
				[
				above = - 1.50 cm of T_n-1_0,
				rectangle,
				rounded corners = 8 pt,
				fill = GreenLighter2!25,
				minimum width = 6 mm,
				minimum height = 95 mm
				]
				( ) { };
				\node
				[
				below = 0.750 cm of T_n-1_0
				]
				( ) {\colorbox{GreenLighter2!75} {$\ket{ GHZ_{ 3 } }$}};
				\node
				[
				above = - 1.50 cm of T_0_m-1,
				rectangle,
				rounded corners = 8 pt,
				fill = RedPurple!25,
				minimum width = 6 mm,
				minimum height = 95 mm
				]
				( ) { };
				\node
				[
				below = 0.750 cm of T_0_m-1
				]
				( ) {\colorbox{RedPurple!75} {$\ket{ GHZ_{ 3 } }$}};
				\node
				[
				above = - 1.50 cm of T_0_0,
				rectangle,
				rounded corners = 8 pt,
				fill = GreenLighter2!25,
				minimum width = 6 mm,
				minimum height = 95 mm
				]
				( ) { };
				\node
				[
				below = 0.750 cm of T_0_0
				]
				( ) {\colorbox{GreenLighter2!75} {$\ket{ GHZ_{ 3 } }$}};
			\end{scope}
			\node
			[
			anchor = east,
			below = 1.00 cm of Trent
			]
			(PhantomNode) { };
		\end{tikzpicture}
		\caption{This figure visualizes the fact that in every triplet of quantum registers $(TIR_{ i }, SIR_{ i }, AIR_{ i })$ the $m$ qubits in the corresponding positions belong to the same $\ket{ GHZ_{ 3 } }$ triplet. To make this point clear, they are drawn with the same color.}
		\label{fig: The Entanglement Distribution Scheme}
	\end{figure}
\end{tcolorbox}

\section{The GHZ$_{ 3 }$MQPEC protocol} \label{sec: The GHZ$_{ 3 }$MQPEC Protocol}

This section is devoted to a thorough presentation of our protocol. The protocol is designed to accomplish quantum private equality comparison among an arbitrary large number $n$ of millionaires using only $\ket{ GHZ_{ 3 } }$ states. From now on we shall refer to it simply by its acronym GHZ$_{ 3 }$MQPEC. Its most important characteristic traits are outlined below.

\begin{itemize}
	[ left = 0.00 cm, labelsep = 0.50 cm ]
	\item	\textbf{Ease of implementation.} Besides EPR pairs, $\ket{ GHZ_{ 3 } }$ triplets are arguably the easiest maximally entangled states that can be produced by modern quantum apparatus. The fact that, irrespective of the number $n$ of millionaires, the GHZ$_{ 3 }$MQPEC protocol can be executed with $\ket{ GHZ_{ 3 } }$ triplets exclusively, guarantees the feasibility of its implementation.
	\item	\textbf{Efficiency \& scalability.} The forthcoming mathematical analysis in section \ref{sec: Efficiency & Security Analysis} corroborates the efficiency of the GHZ$_{ 3 }$MQPEC protocol. Moreover, one of its most important qualitative features, is that it the required quantum resources scale \emph{linearly} both in the number $n$ of millionaires and the number $m$ of qubits required to store their fortunes, which further enhances its potential applicability.
	\item	\textbf{Parallel or sequential execution.} Although the execution of the quantum part of the GHZ$_{ 3 }$MQPEC protocol is envisioned to take place completely in parallel, it is also possible to be implemented sequentially. So, if the quantum resources do not suffice for the execution of the protocol in one go, it is possible to partition the $n$ of millionaires into smaller groups and process these groups sequentially, even to the point of executing the protocol serially for one millionaire at a time.
\end{itemize}

\begin{tcolorbox}
	[
	enhanced,
	breakable,
	grow to left by = 0.00 cm,
	grow to right by = 0.00 cm,
	colback = WordAquaLighter80,			
	enhanced jigsaw,						
	sharp corners,
	toprule = 0.01 pt,
	bottomrule = 0.01 pt,
	leftrule = 0.1 pt,
	rightrule = 0.1 pt,
	sharp corners,
	center title,
	fonttitle = \bfseries
	]
	\begin{center}
		{ \large \textbf{ The intuition behind the} GHZ$_{ 3 }$MQPEC \textbf{protocol } }
	\end{center}
	
	The GHZ$_{ 3 }$MQPEC protocol is based on entanglement. It is the phenomenon of entanglement that allows the secret and untraceable embedding of information in the global state of the quantum system state. The precise mathematical formulation of the correlations among the individual subsystem as encoded into the global state, takes the form given by equation \eqref{eq: Hadamard Entanglement Property}, which we call the \textbf{Hadamard Entanglement Property}.
	
	The other prevalent idea is that instead of having Alice$_{ i }$ and Alice$_{ j }$ directly comparing their fortunes $\mathbf{ f }_{ i }$ and $\mathbf{ f }_{ j }$, $0 \leq j \neq i \leq n - 1$, they can both compare their fortune with a fixed, but secret number $\mathbf{ s }$. If $\mathbf{ s } \oplus \mathbf{ f }_{ i } = \mathbf{ s } \oplus \mathbf{ f }_{ j }$, then $\mathbf{ f }_{ i } = \mathbf{ f }_{ j }$, whereas if $\mathbf{ s } \oplus \mathbf{ f }_{ i } \neq \mathbf{ s } \oplus \mathbf{ f }_{ j }$, then $\mathbf{ f }_{ i } \neq \mathbf{ f }_{ j }$. Sophia is responsible for randomly choosing a secret number $\mathbf{ s }$ and embedding it into the global entanglement. It goes without saying that Sophia must never reveal her secret number $\mathbf{ s }$.
\end{tcolorbox}

\begin{tcolorbox}
	[
	enhanced,
	breakable,
	grow to left by = 0.00 cm,
	grow to right by = 0.00 cm,
	colback = MagentaVeryLight!40,			
	enhanced jigsaw,						
	sharp corners,
	toprule = 0.01 pt,
	bottomrule = 0.01 pt,
	leftrule = 0.1 pt,
	rightrule = 0.1 pt,
	sharp corners,
	center title,
	fonttitle = \bfseries
	]
	\begin{center}
		{ \large \textbf{ The special case of $2$ millionaires } }
	\end{center}
	
	To compare the fortunes of $n > 2$ millionaires, Sophia serves as a secret, albeit fixed point of reference. To give a geometric analogy, one could say that Sophia is the center of a collection of virtual concentric circles and those millionaires that possess the same fortune are points of the same circle. Thus, for $n > 2$ Sophia is indispensable.
	
	Nonetheless, in the very special case where $n = 2$, that is when there are only $2$ millionaires, Sophia becomes unnecessary. In this case, Trent alone suffices. We explain how the protocol and the corresponding quantum circuit simplify in this special case in section \ref{sec: Discussion and Conclusions}.
\end{tcolorbox}

The quantum circuit that implements the GHZ$_{ 3 }$MQPEC protocol for Alice$_{ i }$ is designated by $QC_{ i }$, $0 \leq i \leq n - 1$, and is visualized in Figure \ref{fig: The Quantum Circuit of the GHZ$_{ 3 }$MQPEC Protocol for Alice_$i$}. Considering the depiction of the quantum circuit $QC_{ i }$, $0 \leq i \leq n - 1$, in Figure \ref{fig: The Quantum Circuit of the GHZ$_{ 3 }$MQPEC Protocol for Alice_$i$}, we note the following conventions.

\begin{itemize}
	[ left = 0.00 cm, labelsep = 0.50 cm ]
	\item	For consistency, all quantum circuits adhere to the Qiskit \cite{Qiskit2024} convention in the ordering of qubits. Qiskit follows the little-endian qubit indexing convention, where the least significant qubit is on the right. Accordingly, we place the least significant qubit at the top of the figure and the most significant at the bottom.
	\item	$QC_{ i }$ is the quantum circuit corresponding to Alice$_{ i }$.
	\item	$TIR_{ i }, SIR_{ i }$, and $AIR_{ i }$ are the input registers, each containing $m$ qubits, in the local private circuits of Trent, Sophia, and Alice$_{ i }$, respectively. Their corresponding qubits are entangled, a fact that is visually indicated by the wavy red line connecting them.
	\item	$SOR_{ i }$ is Sophia's single-qubit output register that is initialized to $\ket{ - }$. Analogously, $AOR_{ i }$ is Alice$_{ i }$'s single-qubit output register also initialized to $\ket{ - }$.
	\item	$U_{ \mathbf{ s } }$ is Sophia's unitary transform, as expressed by equation \eqref{eq: Explicit Sophia's Unitary Transform U_s}. $U_{ \mathbf{ f }_{ i } }$ is the unitary transform used by Alice$_{ i }$, which is described by equation \eqref{eq: Explicit Alice_i's Unitary Transform U_f_i}.
	\item	$H^{ \otimes m }$ is the $m$-fold Hadamard transform.
	\item	Alice$_{ i }$, $0 \leq i \leq n - 1$, Sophia, and Trent act privately and secretly on their own quantum circuits. Alice$_{ i }$'s private circuit is structurally similar to Sophia's, but there is a critical difference. Alice$_{ i }$ encodes her fortune $\mathbf{ f }_{ i }$ into the entangled state of the system via her unitary transform $U_{ \mathbf{ f }_{ i } }$. Similarly, Sophia embeds her secret number $\mathbf{ s }$ into the global system through her unitary transform $U_{ \mathbf{ s } }$. Since, in general, $\mathbf{ f }_{ i } \neq \mathbf{ s }$, the unitary transforms $U_{ \mathbf{ f }_{ i } }$ and $U_{ \mathbf{ s } }$ are different.
\end{itemize}

\begin{tcolorbox}
	[
	enhanced,
	breakable,
	grow to left by = 0.00 cm,
	grow to right by = 0.00 cm,
	colback = WordVeryLightTeal!25,			
	enhanced jigsaw,						
	sharp corners,
	toprule = 1.0 pt,
	bottomrule = 1.0 pt,
	leftrule = 0.1 pt,
	rightrule = 0.1 pt,
	sharp corners,
	center title,
	fonttitle = \bfseries
	]
	\begin{figure}[H]
		\centering
		\begin{tikzpicture}[ scale = 0.900, transform shape ]
			\begin{yquant}
				nobit AUX_A_i_0;
				[ name = Alice_i ] qubits { $AIR_{ i }$ } AIR_i;
				qubit { $AOR_{ i }$: \ $\ket{ - }$ } AOR_i;
				nobit AUX_A_i_1;
				[ name = spaceAS, register/minimum height = 8 mm ] nobit spaceAS;
				nobit AUX_S_0;
				[ name = Sophia ] qubits { $SIR_{ i }$ } SIR_i;
				qubit { $SOR_{ i }$: \ $\ket{ - }$ } SOR_i;
				nobit AUX_S_1;
				nobit AUX_S_2;
				[ name = spaceST, register/minimum height = 8 mm ] nobit spaceST;
				nobit AUX_T_i;
				[ name = Trent ] qubits { $TIR_{ i }$ } TIR_i;
				nobit AUX_T_1;
				nobit AUX_T_2;
				[ name = Ph0, WordBlueDarker, line width = 0.50 mm, label = { [ label distance = 0.20 cm ] north: Initial State } ]
				barrier ( - ) ;
				[ draw = WordGoldAccent1Lighter40, fill = MagentaLight!, radius = 0.7 cm ] box {\color{white} \Large \sf{U}$_{ \mathbf{ f }_{ i } }$} (AIR_i - AOR_i);
				[ draw = WordGoldAccent1Lighter40, fill = MagentaLight, radius = 0.7 cm ] box {\color{white} \Large \sf{U}$_{ \mathbf{ s } }$} (SIR_i - SOR_i);
				[ name = Ph1, WordBlueDarker, line width = 0.50 mm, label = { [ label distance = 0.20 cm ] north: Phase 1 } ]
				barrier ( - ) ;
				[ draw = gray, fill = gray, radius = 0.6 cm ] box {\color{white} \Large \sf{H}$^{ \otimes n }$} AIR_i;
				[ draw = gray, fill = gray, radius = 0.6 cm ] box {\color{white} \Large \sf{H}$^{ \otimes n }$} SIR_i;
				[ draw = gray, fill = gray, radius = 0.6 cm ] box {\color{white} \Large \sf{H}$^{ \otimes n }$} TIR_i;
				[ name = Ph2, WordBlueDarker, line width = 0.50 mm, label = { [ label distance = 0.20 cm ] north: Phase 2 } ]
				barrier ( - ) ;
				[ line width = .250 mm, draw = white, fill = gray, radius = 0.6 cm ] measure AIR_i;
				[ line width = .250 mm, draw = white, fill = gray, radius = 0.6 cm ] measure SIR_i;
				[ line width = .250 mm, draw = white, fill = gray, radius = 0.6 cm ] measure TIR_i;
				[ name = Ph3, WordBlueDarker, line width = 0.50 mm, label = { [ label distance = 0.20 cm ] north: Measurement } ]
				barrier ( - ) ;
				output { $\ket{ \mathbf{ y }_{ 0 } }$ } AIR_i;
				output { $\ket{ \mathbf{ y }_{ 1 } }$ } SIR_i;
				output { $\ket{ \mathbf{ y }_{ 2 } }$ } TIR_i;
				\node [ below = 5.00 cm ] at (Ph0) { $\ket{ \psi_{ 0 } }$ };
				\node [ below = 5.00 cm ] at (Ph1) { $\ket{ \psi_{ 1 } }$ };
				\node [ below = 5.00 cm ] at (Ph2) { $\ket{ \psi_{ 2 } }$ };
				\node [ below = 5.00 cm ] at (Ph3) { $\ket{ \psi_{ f } }$ };
				\node
				[
				charlie,
				female,
				scale = 1.50,
				anchor = center,
				left = 0.50 cm of Alice_i,
				label = { [ label distance = 0.00 cm ] west: Alice$_{ i }$ }
				]
				() { };
				\node
				[
				alice,
				scale = 1.50,
				anchor = center,
				left = 0.50 cm of Sophia,
				label = { [ label distance = 0.00 cm ] west: Sophia }
				]
				() { };
				\node
				[
				maninblack,
				scale = 1.50,
				anchor = center,
				left = 0.50 cm of Trent,
				label = { [ label distance = 0.00 cm ] west: Trent }
				]
				() { };
				\begin{scope} [ on background layer ]
					\node [ above right = - 0.30 cm and 0.70 cm of spaceAS, rectangle, fill = WordAquaLighter60, text width = 10.00 cm, align = center, minimum height = 10 mm ] { \bf Possibly Spatially Separated };
					\node [ above right = - 0.30 cm and 0.70 cm of spaceST, rectangle, fill = WordAquaLighter60, text width = 10.00 cm, align = center, minimum height = 10 mm ] { \bf Possibly Spatially Separated };
				\end{scope}
			\end{yquant}
			\node
			[
			above right = 3.00 cm and 5.750 cm of Alice_i,
			anchor = center,
			shade,
			top color = GreenTeal, bottom color = black,
			rectangle,
			text width = 11.00 cm,
			align = center
			]
			(Label)
			{ \color{white}
				The quantum circuit $QC_{ i }$ for Alice$_{ i }$
				\\
			};
			\node [ anchor = center, below = 1.00 cm of Trent ] (PhantomNode) { };
			\scoped [ on background layer ]
			\draw
			[ RedPurple, -, >=stealth, line width = 0.75 mm, decoration = coil, decorate ]
			( $ (Trent.east) + ( 0.5 mm, 0 mm ) $ ) node [ circle, fill, minimum size = 1.5 mm ] () {} -- ( $ (Sophia.east) + ( 0.5 mm, 0 mm ) $ ) node [ circle, fill, minimum size = 1.5 mm ] () {} -- ( $ (Alice_i.east) + ( 0.5 mm, 0 mm ) $ ) node [ circle, fill, minimum size = 1.5 mm ] () {};
		\end{tikzpicture}
		\caption{The above circuit $QC_{ i }$, $0 \leq i \leq n - 1$, embeds Alice$_{ i }$'s fortune $\mathbf{ f }_{ i }$ and Sophia's secret number $\mathbf{ s }$ into the global state of the system. The potentially spatially separated private circuits operated by Alice$_{ i }$, Sophia, and Trent, are linked due to entanglement, indicated by the wavy red line connecting $TIR_{ i }, SIR_{ i }$ and $AIR_{ i }$, and form one composite system.}
		\label{fig: The Quantum Circuit of the GHZ$_{ 3 }$MQPEC Protocol for Alice_$i$}
	\end{figure}
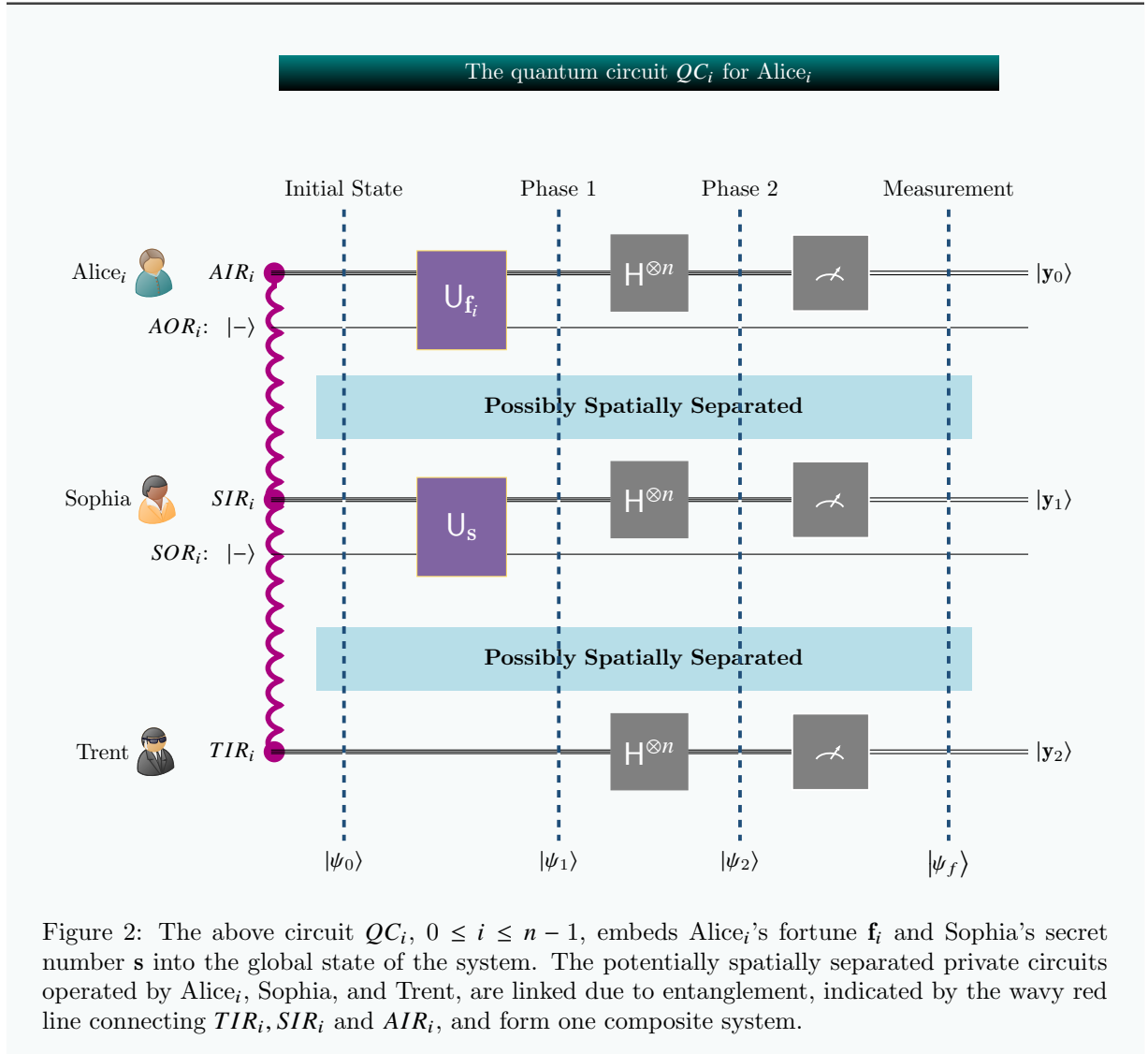
\end{tcolorbox}

\begin{tcolorbox}
	[
	enhanced,
	breakable,
	grow to left by = 0.00 cm,
	grow to right by = 0.00 cm,
	colback = MagentaLight!03,			
	enhanced jigsaw,					
	sharp corners,
	toprule = 1.0 pt,
	bottomrule = 1.0 pt,
	leftrule = 0.1 pt,
	rightrule = 0.1 pt,
	sharp corners,
	center title,
	fonttitle = \bfseries
	]
	\begin{figure}[H]
		\centering
		\begin{tikzpicture} [ scale = 0.750, transform shape ]
			\node
			[
			draw,
			fill = WordBlueDarker50,
			text = white,
			align = center,
			minimum width = 2.00 cm,
			minimum height = 2.00 cm,
			]
			(Alice 0 Quantum Circuit)
			{
				{ \large $QC_{ 0 }$ }
				\\
				{ \small Alice$_{ 0 }$ }
			};
			\node
			[
			below = 0.50 of {Alice 0 Quantum Circuit},
			align = center,
			]
			(Alice 0-i Dots) { \LARGE \dots };
			\node
			[
			below = 0.50 of {Alice 0-i Dots},
			draw,
			fill = WordBlueDarker50,
			text = white,
			align = center,
			minimum width = 2.00 cm,
			minimum height = 2.00 cm,
			]
			(Alice i Quantum Circuit)
			{
				{ \large $QC_{ i }$ }
				\\
				{ \small Alice$_{ i }$ }
			};
			\node
			[
			below = 0.50 of {Alice i Quantum Circuit},
			align = center,
			]
			(Alice i-n-1 Dots) { \LARGE \dots };
			\node
			[
			below = 0.50 of {Alice i-n-1 Dots},
			draw,
			fill = WordBlueDarker50,
			text = white,
			align = center,
			minimum width = 2.00 cm,
			minimum height = 2.00 cm,
			]
			(Alice n-1 Quantum Circuit)
			{
				{ \large $QC_{ n - 1 }$ }
				\\
				{ \small Alice$_{ n - 1 }$ }
			};
			\node
			[
			right = 3.00 of {Alice 0 Quantum Circuit},
			shade,
			top color = WordAquaDarker25, bottom color = black,
			text = white,
			align = center,
			minimum width = 1.750 cm,
			minimum height = 1.00 cm,
			]
			(Alice 0 to Trent) { $\mathbf{ s } \oplus \mathbf{ f }_{ 0 }$ };
			\node
			[
			right = 3.00 of {Alice i Quantum Circuit},
			shade,
			top color = WordAquaDarker25, bottom color = black,
			text = white,
			align = center,
			minimum width = 1.750 cm,
			minimum height = 1.00 cm,
			]
			(Alice i to Trent) { $\mathbf{ s } \oplus \mathbf{ f }_{ i }$ };
			\node
			[
			right = 3.00 of {Alice n-1 Quantum Circuit},
			shade,
			top color = WordAquaDarker25, bottom color = black,
			text = white,
			align = center,
			minimum width = 1.750 cm,
			minimum height = 1.00 cm,
			]
			(Alice n-1 to Trent) { $\mathbf{ s } \oplus \mathbf{ f }_{ n - 1 }$ };
			\draw [ WordLightTeal, -stealth, line width = 3.0 mm ] (Alice 0 Quantum Circuit) -- (Alice 0 to Trent);
			\draw [ WordLightTeal, -stealth, line width = 3.0 mm ] (Alice i Quantum Circuit) -- (Alice i to Trent);
			\draw [ WordLightTeal, -stealth, line width = 3.0 mm ] (Alice n-1 Quantum Circuit) -- (Alice n-1 to Trent);
			\node (Quantum Part)
			[
			draw = WordBlueDarker50, line width = 1.5pt, dash pattern = on 1pt off 4pt on 6pt off 4pt, inner sep = 4mm, rectangle, rounded corners, fit = (Alice 0 Quantum Circuit) (Alice n-1 Quantum Circuit), label = { [ label distance = 0.10 cm ] north: \color{Purple} \textbf{ \large Quantum Part } }
			]
			{};
			\node (Classical Part)
			[
			draw = WordBlueDarker50, line width = 1.5pt, dash pattern = on 1pt off 4pt on 6pt off 4pt, inner sep = 4mm, rectangle, rounded corners, fit = (Alice 0 to Trent) (Alice n-1 to Trent), label = { [ label distance = 0.63 cm ] north: \color{GreenLighter2} \textbf{ \large Classical Part } }
			]
			{};
			\node (Parallel Execution)
			[
			draw = WordBlueDarker50,
			line width = 1.0pt,
			inner sep = 8 mm,
			rectangle,
			fit = (Quantum Part) (Classical Part),
			label = { [ label distance = 0.10 cm ] north: \textbf{ \large Parallel Execution } }
			]
			{};
		\end{tikzpicture}
		\caption{In the ideal scenario, the GHZ$_{ 3 }$MQPEC protocol can be executed entirely in parallel, by employing a parallel array of $n$ circuits $QC_{ 0 }, \dots, QC_{ n - 1 }$.}
		\label{fig: The Complete Parallel Array of $n$ Quantum Circuit}
	\end{figure}
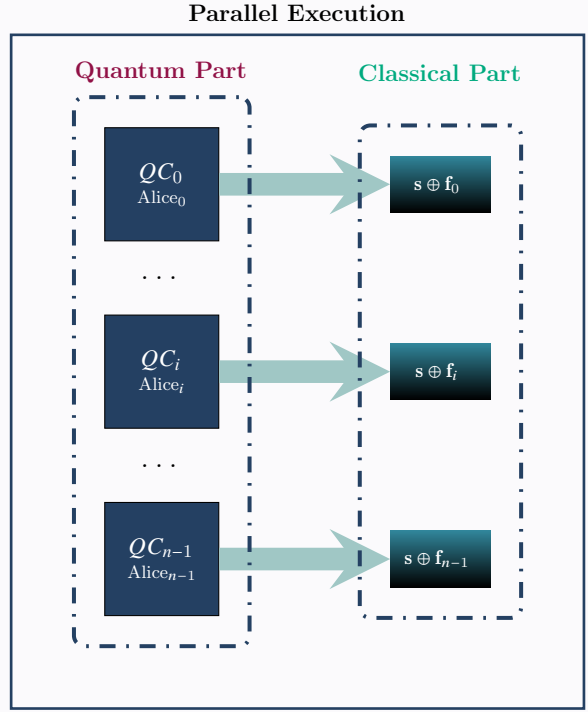
\end{tcolorbox}

\begin{tcolorbox}
	[
	enhanced,
	breakable,
	grow to left by = 0.00 cm,
	grow to right by = 0.00 cm,
	colback = MagentaLight!03,			
	enhanced jigsaw,					
	sharp corners,
	toprule = 1.0 pt,
	bottomrule = 1.0 pt,
	leftrule = 0.1 pt,
	rightrule = 0.1 pt,
	sharp corners,
	center title,
	fonttitle = \bfseries
	]
	\begin{figure}[H]
		\centering
		\begin{tikzpicture} [ scale = 0.750, transform shape ]
			\node
			[
			draw,
			fill = WordBlueDarker50,
			text = white,
			align = center,
			minimum width = 1.250 cm,
			minimum height = 1.250 cm,
			]
			(Alice i Quantum Circuit)
			{
				{ $QC_{ i }$ }
				\\
				{ \footnotesize Alice$_{ i }$ }
			};
			\node
			[
			right = 0.25 of {Alice i Quantum Circuit},
			align = center,
			]
			(Alice i-0 Dots) { \large \dots };
			\node
			[
			right = 0.25 of {Alice i-0 Dots},
			draw,
			fill = WordBlueDarker50,
			text = white,
			align = center,
			minimum width = 1.250 cm,
			minimum height = 1.250 cm,
			]
			(Alice 0 Quantum Circuit)
			{
				{ $QC_{ 0 }$ }
				\\
				{ \footnotesize Alice$_{ 0 }$ }
			};
			\node
			[
			right = 1.25 of {Alice 0 Quantum Circuit},
			draw,
			fill = WordBlueDarker50,
			text = white,
			align = center,
			minimum width = 1.250 cm,
			minimum height = 1.250 cm,
			]
			(Alice j Quantum Circuit)
			{
				{ $QC_{ j }$ }
				\\
				{ \footnotesize Alice$_{ j }$ }
			};
			\node
			[
			right = 0.25 of {Alice j Quantum Circuit},
			align = center,
			]
			(Alice j-i+1 Dots) { \large \dots };
			\node
			[
			right = 0.25 of {Alice j-i+1 Dots},
			draw,
			fill = WordBlueDarker50,
			text = white,
			align = center,
			minimum width = 1.250 cm,
			minimum height = 1.250 cm,
			]
			(Alice i+1 Quantum Circuit)
			{
				{ $QC_{ i + 1 }$ }
				\\
				{ \footnotesize Alice$_{ i + 1 }$ }
			};
			\node
			[
			right = 1.25 of {Alice i+1 Quantum Circuit},
			draw,
			fill = WordBlueDarker50,
			text = white,
			align = center,
			minimum width = 1.250 cm,
			minimum height = 1.250 cm,
			]
			(Alice n-1 Quantum Circuit)
			{
				{ $QC_{ n - 1 }$ }
				\\
				{ \footnotesize Alice$_{ n - 1 }$ }
			};
			\node
			[
			right = 0.25 of {Alice n-1 Quantum Circuit},
			align = center,
			]
			(Alice n-1-j+1 Dots) { \large \dots };
			\node
			[
			right = 0.25 of {Alice n-1-j+1 Dots},
			draw,
			fill = WordBlueDarker50,
			text = white,
			align = center,
			minimum width = 1.250 cm,
			minimum height = 1.250 cm,
			]
			(Alice j+1 Quantum Circuit)
			{
				{ $QC_{ j + 1 }$ }
				\\
				{ \footnotesize Alice$_{ j + 1 }$ }
			};
			\node
			[
			below = 2.50 of {Alice i Quantum Circuit},
			shade,
			top color = WordAquaDarker25, bottom color = black,
			text = white,
			align = center,
			minimum width = 1.250 cm,
			minimum height = 1.00 cm,
			]
			(Alice i to Trent) { \small $\mathbf{ s } \oplus \mathbf{ f }_{ i }$ };
			\node
			[
			below = 2.50 of {Alice 0 Quantum Circuit},
			shade,
			top color = WordAquaDarker25, bottom color = black,
			text = white,
			align = center,
			minimum width = 1.250 cm,
			minimum height = 1.00 cm,
			]
			(Alice 0 to Trent) { \small $\mathbf{ s } \oplus \mathbf{ f }_{ 0 }$ };
			\node
			[
			below = 2.50 of {Alice j Quantum Circuit},
			shade,
			top color = WordAquaDarker25, bottom color = black,
			text = white,
			align = center,
			minimum width = 1.250 cm,
			minimum height = 1.00 cm,
			]
			(Alice j to Trent) { \small $\mathbf{ s } \oplus \mathbf{ f }_{ j }$ };
			\node
			[
			below = 2.50 of {Alice i+1 Quantum Circuit},
			shade,
			top color = WordAquaDarker25, bottom color = black,
			text = white,
			align = center,
			minimum width = 1.250 cm,
			minimum height = 1.00 cm,
			]
			(Alice i+1 to Trent) { \small $\mathbf{ s } \oplus \mathbf{ f }_{ i + 1 }$ };
			\node
			[
			below = 2.50 of {Alice n-1 Quantum Circuit},
			shade,
			top color = WordAquaDarker25, bottom color = black,
			text = white,
			align = center,
			minimum width = 1.250 cm,
			minimum height = 1.00 cm,
			]
			(Alice n-1 to Trent) { \small $\mathbf{ s } \oplus \mathbf{ f }_{ n - 1 }$ };
			\node
			[
			below = 2.50 of {Alice j+1 Quantum Circuit},
			shade,
			top color = WordAquaDarker25, bottom color = black,
			text = white,
			align = center,
			minimum width = 1.250 cm,
			minimum height = 1.00 cm,
			]
			(Alice j+1 to Trent) { \small $\mathbf{ s } \oplus \mathbf{ f }_{ j + 1 }$ };
			\draw [ WordLightTeal, -stealth, line width = 3.0 mm ] (Alice i Quantum Circuit) -- (Alice i to Trent);
			\draw [ WordLightTeal, -stealth, line width = 3.0 mm ] (Alice 0 Quantum Circuit) -- (Alice 0 to Trent);
			\draw [ WordLightTeal, -stealth, line width = 3.0 mm ] (Alice j Quantum Circuit) -- (Alice j to Trent);
			\draw [ WordLightTeal, -stealth, line width = 3.0 mm ] (Alice i+1 Quantum Circuit) -- (Alice i+1 to Trent);
			\draw [ WordLightTeal, -stealth, line width = 3.0 mm ] (Alice n-1 Quantum Circuit) -- (Alice n-1 to Trent);
			\draw [ WordLightTeal, -stealth, line width = 3.0 mm ] (Alice j+1 Quantum Circuit) -- (Alice j+1 to Trent);
			\node (Batch 0)
			[
			draw = WordBlueDarker50,
			line width = 1.0pt,
			inner sep = 4mm,
			rectangle,
			fit = (Alice i Quantum Circuit) (Alice 0 Quantum Circuit),
			label = { [ label distance = - 0.05 cm ] north: \textbf{ Batch 0 } },
			]
			{};
			\node (Batch 00)
			[
			draw = WordBlueDarker50,
			line width = 1.0pt,
			inner sep = 4mm,
			rectangle,
			fit = (Alice i to Trent) (Alice 0 to Trent),
			label = { [ label distance = - 0.05 cm ] north: \textbf{ Batch 0 } },
			]
			{};
			\node (Batch 1)
			[
			draw = WordBlueDarker50,
			line width = 1.0pt,
			inner sep = 4mm,
			rectangle,
			fit = (Alice j Quantum Circuit) (Alice i+1 Quantum Circuit),
			label = { [ label distance = - 0.05 cm ] north: \textbf{ Batch 1 } },
			]
			{};
			\node (Batch 01)
			[
			draw = WordBlueDarker50,
			line width = 1.0pt,
			inner sep = 4mm,
			rectangle,
			fit = (Alice j to Trent) (Alice i+1 to Trent),
			label = { [ label distance = - 0.05 cm ] north: \textbf{ Batch 1 } },
			]
			{};
			\node (Batch 2)
			[
			draw = WordBlueDarker50,
			line width = 1.0pt,
			inner sep = 4mm,
			rectangle,
			fit = (Alice n-1 Quantum Circuit) (Alice j+1 Quantum Circuit),
			label = { [ label distance = - 0.05 cm ] north: \textbf{ Batch 2 } },
			]
			{};
			\node (Batch 02)
			[
			draw = WordBlueDarker50,
			line width = 1.0pt,
			inner sep = 4mm,
			rectangle,
			fit = (Alice n-1 to Trent) (Alice j+1 to Trent),
			label = { [ label distance = - 0.05 cm ] north: \textbf{ Batch 2 } },
			]
			{};
			\draw
			[
			RedPurple,
			->,
			>=stealth,
			line width = 5.0 mm,
			]
			( $ (Batch 0.north) + ( 0.0 mm, 30 mm ) $ )
			--
			( $ (Batch 2.north) + ( 0.0 mm, 30 mm ) $ )
			node [ midway, text = white ] {Time};
			\node (Quantum Part)
			[
			draw = WordBlueDarker50, line width = 1.5pt, dash pattern = on 1pt off 4pt on 6pt off 4pt, inner sep = 4.50 mm, rectangle, rounded corners, fit = (Batch 0) (Batch 2), label = { [ label distance = - 0.01 cm ] north: \color{Purple} \textbf{ Quantum Part } }
			]
			{};
			\node (Classical Part)
			[
			draw = WordBlueDarker50, line width = 1.5pt, dash pattern = on 1pt off 4pt on 6pt off 4pt, inner sep = 4.50 mm, rectangle, rounded corners, fit = (Batch 00) (Batch 02), label = { [ label distance = 0.01 cm ] south: \color{GreenLighter2} \textbf{ Classical Part } }
			]
			{};
			\node (Sequential Execution)
			[
			draw = WordBlueDarker50,
			line width = 1.0pt,
			inner sep = 5 mm,
			rectangle,
			fit = (Quantum Part) (Classical Part),
			label = { [ label distance = 0.07 cm ] north: \textbf{ Sequential Execution } }
			]
			{};
		\end{tikzpicture}
		\caption{If the lack of resources precludes the parallel implementation of the protocol, the $n$ circuits $QC_{ 0 }, \dots, QC_{ n - 1 }$ can be partitioned in sequential batches. In this example, $QC_{ 0 }, \dots, QC_{ n - 1 }$ are partitioned into $3$ batches. The circuits within each batch are employed simultaneously, but the $3$ batches are executed sequentially.}
		\label{fig: The Sequential Execution of $n$ Quantum Circuit in Batches}
	\end{figure}
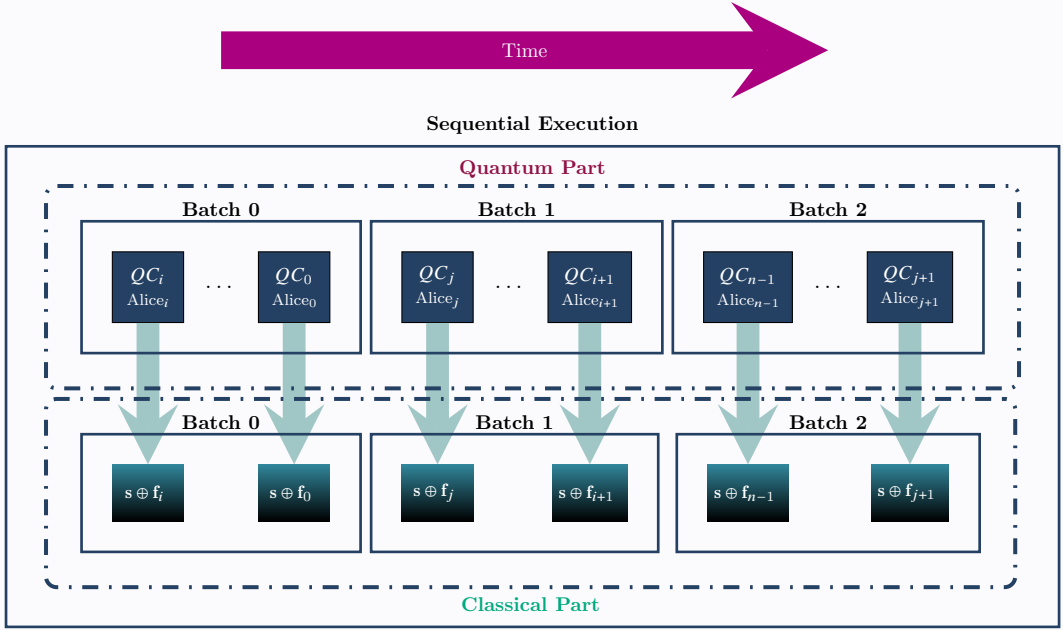
\end{tcolorbox}

Ideally, assuming sufficient quantum resources, we shall employ a parallel array of $n$ such circuits $QC_{ 0 }, \dots, QC_{ n - 1 }$, as visualized in Figure \ref{fig: The Complete Parallel Array of $n$ Quantum Circuit}. In case this is not feasible, due to lack of adequate resources, it is possible to employ these circuits in sequential batches, as we show in Figure \ref{fig: The Sequential Execution of $n$ Quantum Circuit in Batches}. All quantum circuits $QC_{ i }$, $0 \leq i \leq n - 1$, are structurally identical, their only possible difference being in unitary transform $U_{ \mathbf{ f }_{ i } }$ that embeds Alice$_{ i }$'s fortune $\mathbf{ f }_{ i }$ into the system. Obviously, if Alice$_{ i }$ and Alice$_{ j }$'s fortunes $\mathbf{ f }_{ i }$ and $\mathbf{ f }_{ j }$, $0 \leq j \neq i \leq n - 1$, are equal, the corresponding unitary transforms $U_{ \mathbf{ f }_{ i } }$ and $U_{ \mathbf{ f }_{ j } }$ are the same; otherwise they are different.

Utilizing \eqref{eq: m-Fold GHZ_{ 3 } Triplets}, we can express the initial state $\ket{ \psi_{ 0 } }$ of the quantum circuit depicted in Figure \ref{fig: The Quantum Circuit of the GHZ$_{ 3 }$MQPEC Protocol for Alice_$i$} as
\begin{align} \label{eq: GHZ$_{ 3 }$MQPEC Protocol Initial State}
	\ket{ \psi_{ 0 } }
	=
	\frac { 1 } { \sqrt{ 2^{ m } } }
	\sum_{ \mathbf{ x } \in \mathbb{ B }^{ m } }
	\
	\ket{ \mathbf{ x } }_{ T }
	\
	\ket{ - }_{ S }
	\
	\ket{ \mathbf{ x } }_{ S }
	\
	\ket{ - }_{ i }
	\
	\ket{ \mathbf{ x } }_{ i }
	\ .
\end{align}
As mentioned previously, throughout our analysis we shall use the subscripts $T$, $S$, and $i$ to make clear that we refer to the kets in the private circuits of Trent, Sophia, and Alice$_{ i }$, respectively. Hopefully, this practice will enhance readability and avoid any ambiguity.

The execution of the protocol commences in earnest by having Sophia and Alice$_{ i }$ act on their private quantum circuits via their secret unitary transforms $U_{ \mathbf{ s } }$ and $U_{ \mathbf{ f }_{ i } }$. By doing so, both of them encode their secret numbers into the entangled state of the composite circuit. Let us emphasize that by acting on their own individual local circuits, Sophia and Alice$_{ i }$ embed their private number into the three entangled input registers of the composite distributed, circuit.

Sophia's unitary transform $U_{ \mathbf{ s } }$ is based on the scheme $U_{ \mathbf{ s } } \colon \ket{ y } \ \ket{ \mathbf{ x } }$ $\rightarrow$ $\ket{ y \oplus \left( \mathbf{ s } \bullet \mathbf{ x } \right) } \ \ket{ \mathbf{ x } }$, which, for the particular circuit $QC_{ i }$, can be explicitly written as
\begin{align}
	U_{ \mathbf{ s } }
	&\colon
	\ket{ - }_{ S }
	\
	\ket{ \mathbf{ x } }_{ S }
	\rightarrow
	( - 1 )^{ \mathbf{ s } \bullet \mathbf{ x } }
	\
	\ket{ - }_{ S }
	\
	\ket{ \mathbf{ x } }_{ S }
	\ .
	\label{eq: Explicit Sophia's Unitary Transform U_s}
\end{align}
The intuition behind Alice$_{ i }$'s unitary transform $U_{ \mathbf{ f }_{ i } }$ is quite similar, since it implements the scheme $U_{ \mathbf{ f }_{ i } } \colon \ket{ y } \ \ket{ \mathbf{ x } }$ $\rightarrow$ $\ket{ y \oplus \left( \mathbf{ f }_{ i } \bullet \mathbf{ x } \right) } \ \ket{ \mathbf{ x } }$, which, for the our circuit $QC_{ i }$, assumes the following explicit form
\begin{align}
	U_{ \mathbf{ f }_{ i } }
	&\colon
	\ket{ - }_{ i }
	\
	\ket{ \mathbf{ x } }_{ i }
	\rightarrow
	( - 1 )^{ \mathbf{ f }_{ i } \bullet \mathbf{ x } }
	\
	\ket{ - }_{ i }
	\
	\ket{ \mathbf{ x } }_{ i }
	\ .
	\label{eq: Explicit Alice_i's Unitary Transform U_f_i}
\end{align}
The cumulative effect of both unitary transforms drives the quantum circuit into the next state $\ket{ \psi_{ 1 } }$.
\begin{align} \label{eq: GHZ$_{ 3 }$MQPEC Protocol Phase 1}
	\ket{ \psi_{ 1 } }
	&=
	\frac { 1 } { \sqrt{ 2^{ m } } }
	\sum_{ \mathbf{ x } \in \mathbb{ B }^{ m } }
	\
	\ket{ \mathbf{ x } }_{ T }
	\
	\left(
	U_{ \mathbf{ s } }
	\
	\ket{ - }_{ S }
	\
	\ket{ \mathbf{ x } }_{ S }
	\right)
	\
	\left(
	U_{ \mathbf{ f }_{ i } }
	\
	\ket{ - }_{ i }
	\
	\ket{ \mathbf{ x } }_{ i }
	\right)
	\nonumber \\
	&\hspace{-0.37 cm} \overset { \eqref{eq: Explicit Sophia's Unitary Transform U_s}, \eqref{eq: Explicit Alice_i's Unitary Transform U_f_i} } { = }
	\frac { 1 } { \sqrt{ 2^{ m } } }
	\sum_{ \mathbf{ x } \in \mathbb{ B }^{ m } }
	\
	\ket{ \mathbf{ x } }_{ T }
	\
	( - 1 )^{ \mathbf{ s } \bullet \mathbf{ x } }
	\
	\ket{ - }_{ S }
	\
	\ket{ \mathbf{x} }_{ S }
	\
	( - 1 )^{ \mathbf{ f }_{ i } \bullet \mathbf{ x } }
	\
	\ket{ - }_{ i }
	\
	\ket{ \mathbf{ x } }_{ i }
	\nonumber \\
	&=
	\frac { 1 } { \sqrt{ 2^{ m } } }
	\sum_{ \mathbf{ x } \in \mathbb{ B }^{ m } }
	\
	( - 1 )^{ ( \mathbf{ s } \oplus \mathbf{ f }_{ i } ) \bullet \mathbf{ x } }
	\
	\ket{ \mathbf{ x } }_{ T }
	\
	\ket{ - }_{ S }
	\
	\ket{ \mathbf{x} }_{ S }
	\
	\ket{ - }_{ i }
	\
	\ket{ \mathbf{ x } }_{ i }
	\ .
\end{align}
Therefore, at the end of Phase 1, Sophia and Alice$_{ i }$ have embedded their private numbers $\mathbf{ s }$ and $\mathbf{ f }_{ i }$, known only to them, in an untraceable way into the entangled state $\ket{ \psi_{ 1 } }$ of the composite quantum circuit. Now, it remains to extract this information, so that it allows Trent to compare the fortunes of the millionaires. To decipher the embedded private information, Trent, Sophia, and Alice$_{ i }$ apply the $m$-fold Hadamard transform to their input registers, as shown in Figure \ref{fig: The Quantum Circuit of the GHZ$_{ 3 }$MQPEC Protocol for Alice_$i$}. Consequently, at the end of Phase 2, the state of the system has become $\ket{ \psi_{ 2 } }$.
\begin{align} \label{eq: GHZ$_{ 3 }$MQPEC Protocol Phase 2 - 1}
	\ket{ \psi_{ 2 } }
	&\hspace{-0.08 cm}\overset { \eqref{eq: GHZ$_{ 3 }$MQPEC Protocol Phase 1} } { = }
	\frac { 1 } { \sqrt{ 2^{ m } } }
	\sum_{ \mathbf{ x } \in \mathbb{ B }^{ m } }
	\
	( - 1 )^{ ( \mathbf{ s } \oplus \mathbf{ f }_{ i } ) \bullet \mathbf{ x } }
	\
	H^{ \otimes m }
	\ket{ \mathbf{ x } }_{ T }
	\
	\ket{ - }_{ S }
	\
	H^{ \otimes m }
	\ket{ \mathbf{x} }_{ S }
	\
	\ket{ - }_{ i }
	\
	H^{ \otimes m }
	\ket{ \mathbf{ x } }_{ i }
	\ .
\end{align}
By invoking relation \eqref{eq: Hadamard m-Fold Ket x}, we may analyze $H^{ \otimes m } \ket{ \mathbf{ x } }_{ T }$, $H^{ \otimes m } \ket{ \mathbf{ x } }_{ S }$, and $H^{ \otimes m } \ket{ \mathbf{ x } }_{ i }$ further.
\begin{align*} 
	H^{ \otimes m }
	\ket{ \mathbf{ x } }_{ T }
	&=
	\frac { 1 } { \sqrt{ 2^{ m } } }
	\sum_{ \mathbf{ z }_{ 2 } \in \mathbb{ B }^{ m } }
	\
	( - 1 )^{ \mathbf{ z }_{ 2 } \bullet \mathbf{ x } }
	\
	\ket{ \mathbf{ z }_{ 2 } }_{ T }
	\\
	H^{ \otimes m }
	\ket{ \mathbf{ x } }_{ S }
	&=
	\frac { 1 } { \sqrt{ 2^{ m } } }
	\sum_{ \mathbf{ z }_{ 1 } \in \mathbb{ B }^{ m } }
	\
	( - 1 )^{ \mathbf{ z }_{ 1 } \bullet \mathbf{ x } }
	\
	\ket{ \mathbf{ z }_{ 1 } }_{ S }
	\\
	H^{ \otimes m }
	\ket{ \mathbf{ x } }_{ i }
	&=
	\frac { 1 } { \sqrt{ 2^{ m } } }
	\sum_{ \mathbf{ z }_{ 0 } \in \mathbb{ B }^{ m } }
	\
	( - 1 )^{ \mathbf{ z }_{ 0 } \bullet \mathbf{ x } }
	\
	\ket{ \mathbf{ z }_{ 0 } }_{ i }
\end{align*}
If we make the above substitutions, $\ket{ \psi_{ 2 } }$ can be cast in an alternative form as shown below.
\begin{align} \label{eq: GHZ$_{ 3 }$MQPEC Protocol Phase 2 - 2}
	\ket{ \psi_{ 2 } }
	=
	\frac { 1 } { 2^{ 2 m } }
	\sum_{ \mathbf{ z }_{ 2 } \in \mathbb{ B }^{ m } }
	\sum_{ \mathbf{ z }_{ 1 } \in \mathbb{ B }^{ m } }
	\sum_{ \mathbf{ z }_{ 0 } \in \mathbb{ B }^{ m } }
	\sum_{ \mathbf{ x } \in \mathbb{ B }^{ m } }
	\
	( - 1 )^{ ( \mathbf{ s } \oplus \mathbf{ f }_{ i } \oplus \mathbf{ z }_{ 2 } \oplus \mathbf{ z }_{ 1 } \oplus \mathbf{ z }_{ 0 } )
		\bullet \mathbf{ x } }
	\
	\ket{ \mathbf{ z }_{ 2 } }_{ T }
	\
	\ket{ - }_{ S }
	\
	\ket{ \mathbf{ z }_{ 1 } }_{ S }
	\
	\ket{ - }_{ i }
	\
	\ket{ \mathbf{ z }_{ 0 } }_{ i }
	\ .
\end{align}
This last formula can be written in a more readable form in view of the characteristic inner product property for zero \eqref{eq: Inner Product Modulo $2$ Property For Zero} and nonzero \eqref{eq: Inner Product Modulo $2$ Property For NonZero} bit vectors. In this specific setting, the application of these properties leads to the following reasoning.
\begin{itemize}
	[ left = 0.00 cm, labelsep = 0.50 cm ]
	\item	
	If $\mathbf{ s } \oplus \mathbf{ f }_{ i } \oplus \mathbf{ z }_{ 2 } \oplus \mathbf{ z }_{ 1 } \oplus \mathbf{ z }_{ 0 } \neq \mathbf{ 0 }$, or, equivalently, $\mathbf{ z }_{ 2 } \oplus \mathbf{ z }_{ 1 } \oplus \mathbf{ z }_{ 0 }$ $\neq$ $\mathbf{ s } \oplus \mathbf{ f }_{ i }$, the sum
	\begin{align*}
		\sum_{ \mathbf{ x } \in \mathbb{ B }^{ m } }
		\
		( - 1 )^{ ( \mathbf{ s } \oplus \mathbf{ f }_{ i } \oplus \mathbf{ z }_{ 2 } \oplus \mathbf{ z }_{ 1 } \oplus \mathbf{ z }_{ 0 } ) \bullet \mathbf{ x } }
		\
		\ket{ \mathbf{ z }_{ 2 } }_{ T }
		\
		\ket{ - }_{ S }
		\
		\ket{ \mathbf{ z }_{ 1 } }_{ S }
		\
		\ket{ - }_{ i }
		\
		\ket{ \mathbf{ z }_{ 0 } }_{ i }
	\end{align*}
	appearing in \eqref{eq: GHZ$_{ 3 }$MQPEC Protocol Phase 2 - 2} becomes just $0$.
	\item	
	If $\mathbf{ s } \oplus \mathbf{ f }_{ i } \oplus \mathbf{ z }_{ 2 } \oplus \mathbf{ z }_{ 1 } \oplus \mathbf{ z }_{ 0 } = \mathbf{ 0 }$, or, equivalently, $\mathbf{ z }_{ 2 } \oplus \mathbf{ z }_{ 1 } \oplus \mathbf{ z }_{ 0 }$ $=$ $\mathbf{ s } \oplus \mathbf{ f }_{ i }$, the sum
	\begin{align*}
		\sum_{ \mathbf{ x } \in \mathbb{ B }^{ m } }
		\
		( - 1 )^{ ( \mathbf{ s } \oplus \mathbf{ f }_{ i } \oplus \mathbf{ z }_{ 2 } \oplus \mathbf{ z }_{ 1 } \oplus \mathbf{ z }_{ 0 } ) \bullet \mathbf{ x } }
		\
		\ket{ \mathbf{ z }_{ 2 } }_{ T }
		\
		\ket{ - }_{ S }
		\
		\ket{ \mathbf{ z }_{ 1 } }_{ S }
		\
		\ket{ - }_{ i }
		\
		\ket{ \mathbf{ z }_{ 0 } }_{ i }
	\end{align*}
	becomes
	\begin{align} \label{eq: Inner Product Modulo $2$ Property for the GHZ$_{ 3 }$MQPEC Protocol}
		2^{ m }
		\
		\ket{ \mathbf{ z }_{ 2 } }_{ T }
		\
		\ket{ - }_{ S }
		\
		\ket{ \mathbf{ z }_{ 1 } }_{ S }
		\
		\ket{ - }_{ i }
		\
		\ket{ \mathbf{ z }_{ 0 } }_{ i }
		\ .
	\end{align}
\end{itemize}
Thus, $\ket{ \psi_{ 2 } }$ can be written in the simpler form given below.
\begin{align} \label{eq: GHZ$_{ 3 }$MQPEC Protocol Phase 2 - 3}
	\ket{ \psi_{ 2 } }
	=
	\frac { 1 } { 2^{ m } }
	\sum_{ \mathbf{ z }_{ 2 } \in \mathbb{ B }^{ m } }
	\sum_{ \mathbf{ z }_{ 1 } \in \mathbb{ B }^{ m } }
	\sum_{ \mathbf{ z }_{ 0 } \in \mathbb{ B }^{ m } }
	\
	\ket{ \mathbf{ z }_{ 2 } }_{ T }
	\
	\ket{ - }_{ S }
	\
	\ket{ \mathbf{ z }_{ 1 } }_{ S }
	\
	\ket{ - }_{ i }
	\
	\ket{ \mathbf{ z }_{ 0 } }_{ i }
	\ ,
\end{align}
where
\begin{align} \label{eq: Hadamard Entanglement Property}
	\mathbf{ z }_{ 2 }
	\oplus
	\mathbf{ z }_{ 1 }
	\oplus
	\mathbf{ z }_{ 0 }
	=
	\mathbf{ s }
	\oplus
	\mathbf{ f }_{ i }
	\ .
\end{align}
Henceforth, we refer to the relation \eqref{eq: Hadamard Entanglement Property} as the \textbf{Hadamard Entanglement Property} that intertwines the input registers of Trent, Sophia and Alice$_{ i }$. This relation is the aftermath of the initial entanglement among all these input registers. At the end of Phase 2, Sophia and Alice$_{ i }$ have encoded their private numbers $\mathbf{ s }$ and $\mathbf{ f }_{ i }$ as the modulo $2$ sum $\mathbf{ s } \oplus \mathbf{ f }_{ i }$ in the global state of the composite quantum circuit, which has caused this constraint on the contents of the input registers.

To complete the quantum part of the GHZ$_{ 3 }$MQPEC protocol, Trent, Sophia and Alice$_{ i }$ measure their input registers in the computational basis. By this action, the state of the composite system collapses to the final state $\ket{ \psi_{ f } }$.
\begin{align}
	\ket{ \psi_{ f } }
	&=
	\ket{ \mathbf{ y }_{ 2 } }_{ T }
	\
	\ket{ - }_{ S }
	\
	\ket{ \mathbf{ y }_{ 1 } }_{ S }
	\
	\ket{ - }_{ i }
	\
	\ket{ \mathbf{ y }_{ 0 } }_{ i }
	\ ,
	\text{ where }
	\label{eq: GHZ$_{ 3 }$MQPEC Protocol Final Measurement}
	\\
	&\mathbf{ y }_{ 2 }
	\oplus
	\mathbf{ y }_{ 1 }
	\oplus
	\mathbf{ y }_{ 0 }
	=
	\mathbf{ s }
	\oplus
	\mathbf{ f }_{ i }
	\ .
	\label{eq: GHZ$_{ 3 }$MQPEC Protocol Final Sum}
\end{align}
We note that both output registers $SOR_{ i }$ and $AOR_{ i }$ remain in their initial state $\ket{ - }$ throughout the execution of the protocol. In that sense, they could be ignored and removed from \eqref{eq: GHZ$_{ 3 }$MQPEC Protocol Final Measurement}. Nonetheless, we have opted to include them for completeness. Let us emphasize that all the above equations hold true no matter whether the private circuits of the players are in the same locality or are spatially distributed because their validity is due to the initial entanglement among the input registers. In the presence of entanglement, the distance among the players is irrelevant.

From now on, the GHZ$_{ 3 }$MQPEC protocol uses only classical authenticated communication channels. The following actions allow Trent to complete the multiparty equality comparison among the $n$ millionaires, and announce his decision in the form of a single bit for every pair of Alice$_{ i }$ and Alice$_{ j }$, $0 \leq i < j \leq n - 1$, namely, YES$_{ i, j }$ if $\mathbf{ f }_{ i } = \mathbf{ f }_{ j }$, or NO$_{ i, j }$ if $\mathbf{ f }_{ i } \neq \mathbf{ f }_{ j }$.
\begin{enumerate}
	\renewcommand\labelenumi{(\textbf{D}$_\theenumi$)}
	\item	Sophia and Alice$_{ i }$ send to Trent the measured contents of their input registers $\mathbf{ y }_{ 1 }$ and $\mathbf{ y }_{ 0 }$, respectively. The communication takes place via authenticated classical channels, which means that the transmitted messages become public knowledge, but cannot be modified by an adversary.
	\item	Trent, as was stipulated in (\textbf{A}$_{ 4 }$) of subsection \ref{subsec: The Actors}, never divulges the measured contents of his input register $\mathbf{ y }_{ 2 }$ to any other in-game player or outside eavesdropper. This implies that only Trent has the complete information to compute the modulo $2$ sum $\mathbf{ s } \oplus \mathbf{ f }_{ i }$ according to \eqref{eq: GHZ$_{ 3 }$MQPEC Protocol Final Sum}.
	\item	The same procedure is repeated for every millionaire Alice$_{ 0 }$, \dots, Alice$_{ n - 1 }$. This enables Trent to calculate the sequence of $n$ modulo $2$ sums $\mathbf{ s } \oplus \mathbf{ f }_{ 0 }$, \dots, $\mathbf{ s } \oplus \mathbf{ f }_{ n - 1 }$. We note that Trent, as was ordained in (\textbf{A}$_{ 4 }$), never reveals the aforementioned sequence to any other in-game player or outside eavesdropper. The following obvious identity
	\begin{align}
		\mathbf{ s }
		\oplus
		\mathbf{ f }_{ i }
		=
		\mathbf{ s }
		\oplus
		\mathbf{ f }_{ j }
		\Leftrightarrow
		\mathbf{ f }_{ i }
		=
		\mathbf{ f }_{ j }
		\label{eq: Module $2$ Sum Equality Identity}
	\end{align}
	is the key to determine whether two fortunes $\mathbf{ f }_{ i }$ and $\mathbf{ f }_{ j }$, $0 \leq i < j \leq n - 1$, are equal or not. Finally, Trent broadcasts through classical channels a sequence of $\frac { n ( n - 1 ) } { 2 }$ of bits that carry the information YES$_{ i, j }$ or NO$_{ i, j }$, $0 \leq i < j \leq n - 1$, depending on whether $\mathbf{ f }_{ i } = \mathbf{ f }_{ j }$, or $\mathbf{ f }_{ i } \neq \mathbf{ f }_{ j }$.
\end{enumerate}
To get a better feeling for the inner workings of the GHZ$_{ 3 }$MQPEC protocol, we present an illustrative example in the next section.

\section{The GHZ$_{ 3 }$MQPEC protocol in action} \label{sec: The GHZ$_{ 3 }$MQPEC Protocol in Action}

This section contains an example that demonstrates and explains the operation of GHZ$_{ 3 }$MQPEC protocol. For practical reasons, this is a small-scale example utilizing $24$ qubits. A bigger example, involving more qubits, would result in unintelligible figures that would obstruct the readability and understanding of the quantum circuit. This example, which faithfully follows the abstract quantum circuit of Figure \ref{fig: The Quantum Circuit of the GHZ$_{ 3 }$MQPEC Protocol for Alice_$i$}, has been implemented in Qiskit \cite{Qiskit2024} and the resulting quantum circuit is shown in Figure \ref{fig: GHZ$_{ 3 }$MQPEC Example for $3$ Millionaires}.

We assume that $n = 3$ millionaires Alice$_{ 0 }$, Alice$_{ 1 }$, and Alice$_{ 2 }$ with private fortunes $\mathbf{ f }_{ 0 } = 11$, $\mathbf{ f }_{ 1 } = 11$, and $\mathbf{ f }_{ 2 } = 01$ want to compare their fortunes in one go with the mediation of Sophia, whose secret number is $\mathbf{ s } = 10$, and Trent. To keep the qubit count low, we have opted for input registers that have $2$ qubits, i.e., $m = 2$. So, we employ a parallel an array of three circuits $QC_{ 0 }$, $QC_{ 1 }$, and $QC_{ 2 }$, corresponding to Alice$_{ 0 }$, Alice$_{ 1 }$, and Alice$_{ 2 }$, respectively. Below we explain the acronyms used in Figure \ref{fig: GHZ$_{ 3 }$MQPEC Example for $3$ Millionaires}.

\begin{tcolorbox}
	[
		enhanced,
		breakable,
		grow to left by = 1.00 cm,
		grow to right by = 0.00 cm,
		colback = white,			
		enhanced jigsaw,			
		sharp corners,
		toprule = 0.1 pt,
		bottomrule = 0.1 pt,
		leftrule = 0.1 pt,
		rightrule = 0.1 pt,
		sharp corners,
		center title,
		fonttitle = \bfseries
	]
	{\small
		\begin{minipage}[b]{0.315 \textwidth}
			$QC_{ 0 }$ consists of the following components:
			\begin{itemize}
				[ left = 0.00 cm, labelsep = 0.20 cm ]
				\item	Alice$_{ 0 }$'s input register \textit{alice's\_0\_IR} containing $2$ qubits.
				\item	Alice$_{ 0 }$'s single-qubit output register \textit{alice's\_0\_OR} initialized at $\ket{ - }$.
				\item	Sophia's input register \textit{sophia's\_0\_IR} containing $2$ qubits.
				\item	Sophia's single-qubit output register \textit{sophia's\_0\_OR} initialized at $\ket{ - }$.
				\item	Trent's input register \textit{trent's\_0\_IR} containing $2$ qubits.
				\item	\textit{trent's\_0\_IR}, \textit{sophia's\_0\_IR}, and \textit{alice's\_0\_IR} are initialized in the entangled state $\ket{ GHZ_{ 3 } }^{ \otimes 2 }$.
				\item	The $3$ classical registers (each containing $2$ bits) used by Alice$_{ 0 }$, Sophia, and Trent to store their measurements, are collectively referred to as the \textit{classicalRegister0} with a total capacity of $6$ bits.
			\end{itemize}
		\end{minipage} 
		\hspace{3 mm}
		\begin{minipage}[b]{0.315 \textwidth}
			$QC_{ 1 }$ consists of the following components:
			\begin{itemize}
				[ left = 0.00 cm, labelsep = 0.20 cm ]
				\item	Alice$_{ 1 }$'s input register \textit{alice's\_1\_IR} containing $2$ qubits.
				\item	Alice$_{ 1 }$'s single-qubit output register \textit{alice's\_1\_OR} initialized at $\ket{ - }$.
				\item	Sophia's input register \textit{sophia's\_1\_IR} containing $2$ qubits.
				\item	Sophia's single-qubit output register \textit{sophia's\_1\_OR} initialized at $\ket{ - }$.
				\item	Trent's input register \textit{trent's\_1\_IR} containing $2$ qubits.
				\item	\textit{trent's\_1\_IR}, \textit{sophia's\_1\_IR}, and \textit{alice's\_1\_IR} are initialized in the entangled state $\ket{ GHZ_{ 3 } }^{ \otimes 2 }$.
				\item	The $3$ classical registers (each containing $2$ bits) used by Alice$_{ 1 }$, Sophia, and Trent to store their measurements, are collectively referred to as the \textit{classicalRegister1} with a total capacity of $6$ bits.
			\end{itemize}
		\end{minipage}
		\hspace{3 mm}
		\begin{minipage}[b]{0.315 \textwidth}
			$QC_{ 2 }$ consists of the following components:
			\begin{itemize}
				[ left = 0.00 cm, labelsep = 0.20 cm ]
				\item	Alice$_{ 2 }$'s input register \textit{alice's\_2\_IR} containing $2$ qubits.
				\item	Alice$_{ 2 }$'s single-qubit output register \textit{alice's\_2\_OR} initialized at $\ket{ - }$.
				\item	Sophia's input register \textit{sophia's\_2\_IR} containing $2$ qubits.
				\item	Sophia's single-qubit output register \textit{sophia's\_2\_OR} initialized at $\ket{ - }$.
				\item	Trent's input register \textit{trent's\_2\_IR} containing $2$ qubits.
				\item	\textit{trent's\_0\_IR}, \textit{sophia's\_2\_IR}, and \textit{alice's\_2\_IR} are initialized in the entangled state $\ket{ GHZ_{ 3 } }^{ \otimes 2 }$.
				\item	The $3$ classical registers (each containing $2$ bits) used by Alice$_{ 2 }$, Sophia, and Trent to store their measurements, are collectively referred to as the \textit{classicalRegister2} with a total capacity of $6$ bits.
			\end{itemize}
		\end{minipage}
	}
\end{tcolorbox}

Alice$_{ 0 }$, Alice$_{ 1 }$, Alice$_{ 2 }$, and Sophia embed their private numbers in the entangled global state of the composite circuit. The embedding is trivially implemented using exclusively CNOT gates. When they measure their input registers, Alice$_{ 0 }$, Alice$_{ 1 }$, Alice$_{ 2 }$, Sophia, and Trent get one of the $2^{ 24 } = 16,777,216$ equiprobable outcomes. For obvious technical limitations, we can't show all these outcomes, since this would result in an unintelligible figure. Hence, we depict only 16 of them in Figure \ref{fig: GHZ$_{ 3 }$MQPEC Example Measurement Outcomes}.

\begin{tcolorbox}
	[
		enhanced,
		breakable,
		grow to left by = 0.00 cm,
		grow to right by = 0.00 cm,
		colback = white,			
		enhanced jigsaw,			
		sharp corners,
		boxrule = 0.01 pt,
		toprule = 0.01 pt,
		bottomrule = 0.01 pt
	]
	\begin{figure}[H]
		\centering
		\includegraphics
		[
			angle = 90,
			height = 21.000 cm,
			width = 17.000 cm,
			trim = {0 0 0cm 0},
			clip
		]
		{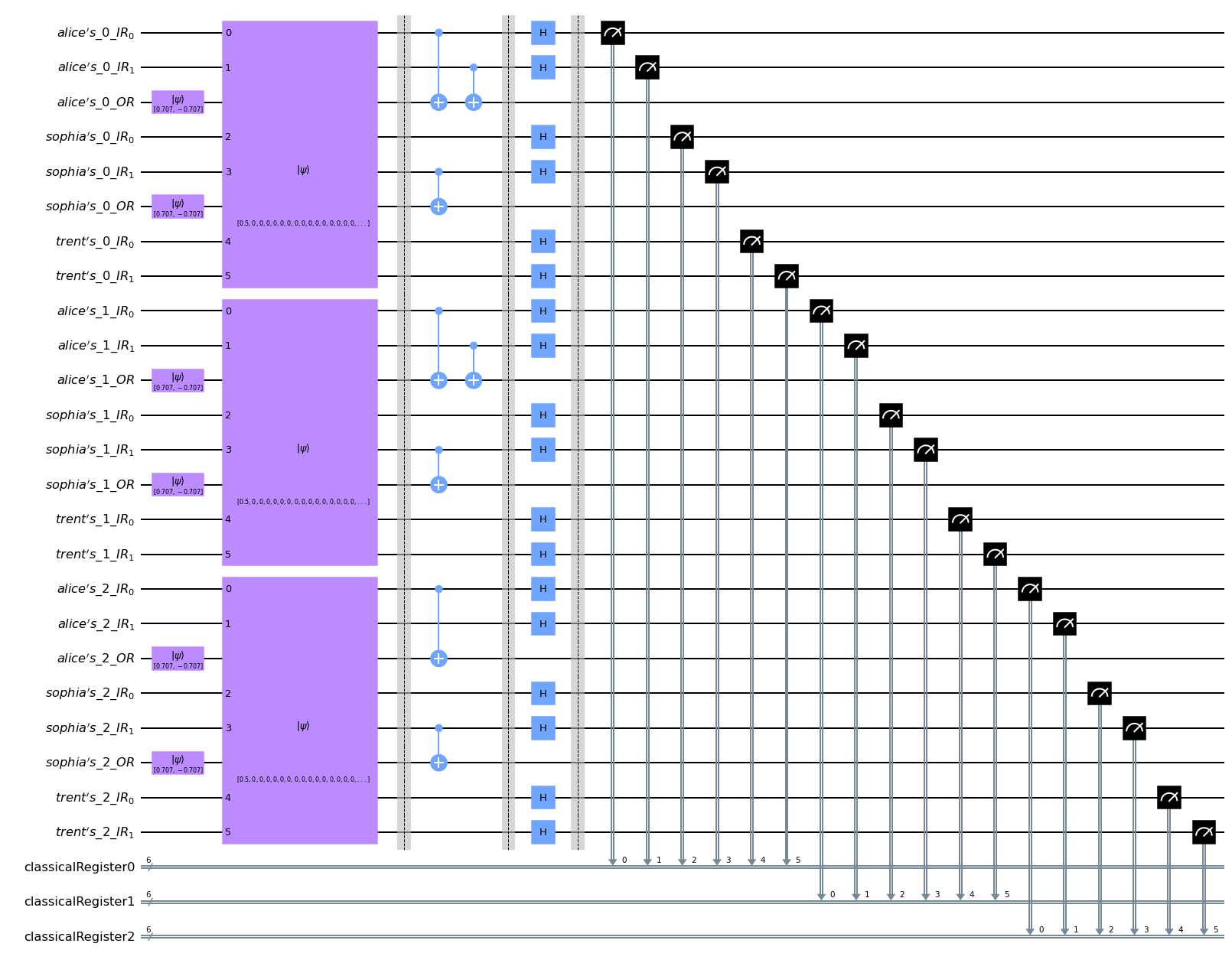}
		\caption{The above quantum circuit, created with Qiskit, implements the GHZ$_{ 3 }$MQPEC protocol for $3$ millionaires, Alice$_{ 0 }$, Alice$_{ 1 }$, and Alice$_{ 2 }$. Their fortunes have been chosen so that they can be represent with $m = 2$ qubits.}
		\label{fig: GHZ$_{ 3 }$MQPEC Example for $3$ Millionaires}
	\end{figure}
\end{tcolorbox}

\begin{tcolorbox}
	[
		enhanced,
		breakable,
		grow to left by = 0.00 cm,
		grow to right by = 0.00 cm,
		colback = white,			
		enhanced jigsaw,			
		sharp corners,
		boxrule = 0.01 pt,
		toprule = 0.01 pt,
		bottomrule = 0.01 pt
	]
	\begin{figure}[H]
		\centering
		\includegraphics
		[
			scale = 0.3,
			angle = 90,
			trim = {0cm 0cm 0cm 0cm},
			clip
		]
		{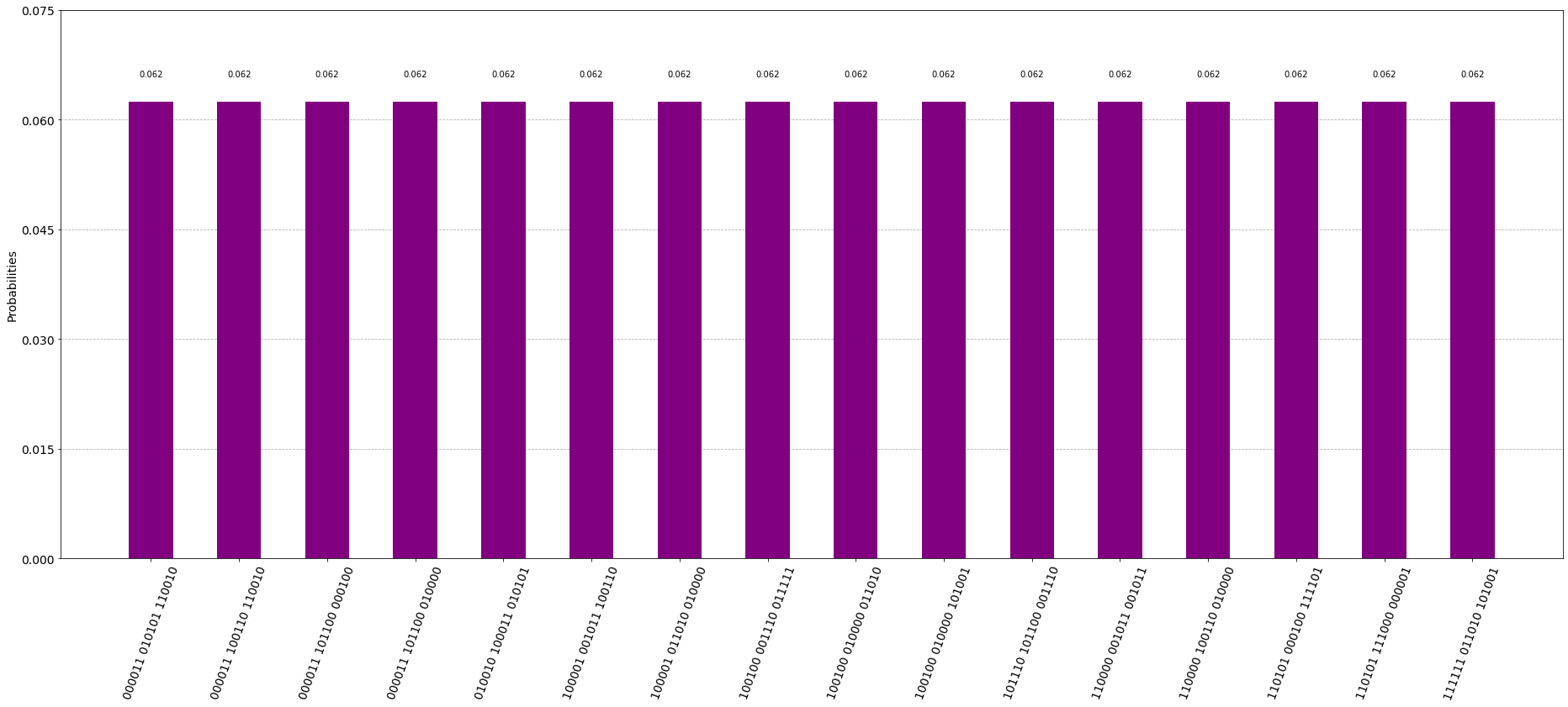}
		\caption{Some of the possible measurements and their corresponding probabilities for the circuit of Figure \ref{fig: GHZ$_{ 3 }$MQPEC Example for $3$ Millionaires}.}
		\label{fig: GHZ$_{ 3 }$MQPEC Example Measurement Outcomes}
	\end{figure}
\end{tcolorbox}

It is straightforward to verify that all outcomes satisfy the Hadamard Entanglement Property, as quantified by \eqref{eq: GHZ$_{ 3 }$MQPEC Protocol Final Sum}. To explain this point beyond any doubt, let us examine the label of the first bar of the histogram of Figure \ref{fig: GHZ$_{ 3 }$MQPEC Example Measurement Outcomes}, which is $\colorbox {WordAquaLighter40} {000011} \ \colorbox {WordAquaLighter60} {010101} \ \colorbox {WordAquaLighter80} {110010}$. The label consists of $3$ parts: $\colorbox {WordAquaLighter40} {000011}$, corresponding to the contents of \textit{classicalRegister2}, $\colorbox {WordAquaLighter60} {010101}$, corresponding to the contents of the \textit{classicalRegister1}, and $\colorbox {WordAquaLighter80} {110010}$, corresponding to the contents of the \textit{classicalRegister0}. This ordering is a direct consequence of the way the quantum circuit in Figure \ref{fig: GHZ$_{ 3 }$MQPEC Example for $3$ Millionaires} was constructed, i.e., the bits of \textit{classicalRegister2} are more significant than those of \textit{classicalRegister1}, which in turn are more significant than those of \textit{classicalRegister0}.

\begin{itemize}
	[ left = 0.00 cm, labelsep = 0.50 cm ]
	\item	\textit{classicalRegister0} is the collective name for the $3$ classical registers (each containing $2$ bits) used by Trent, Sophia, and Alice$_{ 0 }$ to store their measurements. Therefore, the contents of $\colorbox {WordAquaLighter80} {110010}$ are analyzed as shown below.
	\begin{align}
		&\phantom{ = }
		\underbrace { \colorbox {MagentaVeryLight} { 1 1 } }_{ \text{Trent} }
		\underbrace { \colorbox {MagentaVeryLight!40!MyLightRed} { 0 0 } }_{ \text{Sophia} }
		\underbrace { \colorbox {MyLightRed} { 10 } }_{ \text{Alice}_{ 0 } }
		\nonumber \\
		&\phantom{ = }
		\hspace{ 0.125 cm }
		\overset { \downarrow } { \colorbox {RedPurple!80} { $\mathbf{ y }_{ 2 }$ } }
		\hspace{ 0.180 cm }
		\overset { \downarrow } { \colorbox {GreenLighter2} { $\mathbf{ y }_{ 1 }$ } }
		\hspace{ 0.100 cm }
		\overset { \downarrow } { \colorbox {WordBlueVeryLight} { $\mathbf{ y }_{ 0 }$ } }
		\label{eq: classicalRegister0 Content Analysis}
	\end{align}
	This means that $\mathbf{ y }_{ 2 } = 11$, $\mathbf{ y }_{ 1 } = 00$, and $\mathbf{ y }_{ 0 } = 10$. Sophia and Alice$_{ 0 }$ send to Trent  $\mathbf{ y }_{ 1 }$ and $\mathbf{ y }_{ 0 }$ via authenticated classical channels. So, Trent has the complete information to compute the modulo $2$ sum of $\mathbf{ y }_{ 2 }$, $\mathbf{ y }_{ 1 }$, and $\mathbf{ y }_{ 0 }$ and conclude that
	\begin{align}
		\mathbf{ s }
		\oplus
		\mathbf{ f }_{ 0 }
		=
		01
		\label{eq: Alice's 0 Fortune Encoded Form}
		\ ,
	\end{align}
	according to \eqref{eq: GHZ$_{ 3 }$MQPEC Protocol Final Sum}.
	\item	Reasoning in exactly the same way for \textit{classicalRegister1} and \textit{classicalRegister2}, it is straightforward to see that Trent arrives to the additional conclusions
	\begin{align}
		\mathbf{ s }
		\oplus
		\mathbf{ f }_{ 1 }
		&=
		01
		\label{eq: Alice's 1 Fortune Encoded Form}
		\ , \ \text{and}
		\\
		\mathbf{ s }
		\oplus
		\mathbf{ f }_{ 2 }
		&=
		11
		\label{eq: Alice's 2 Fortune Encoded Form}
		\ .
	\end{align}
\end{itemize}

Ergo, Trent ascertains that Alice$_{ 0 }$ and Alice$_{ 1 }$'s fortunes are equal, but different from Alice$_{ 2 }$'s fortune. We emphasize that Sophia, as stipulated in (\textbf{A}$_{ 3 }$), never reveals her secret number $\mathbf{ s }$, so Trent cannot infer $\mathbf{ f }_{ i }$ from $\mathbf{ s } \oplus \mathbf{ f }_{ i }$. Likewise, Trent never divulges $\mathbf{ y }_{ 2 }$ or the modulo $2$ sum $\mathbf{ y }_{ 2 } \oplus \mathbf{ y }_{ 1 } \oplus \mathbf{ y }_{ 0 }$ to any other in-game player or outside eavesdropper, as ordained by (\textbf{A}$_{ 4 }$) and (\textbf{D}$_{ 3 }$). Thus, no entity can gain any information about the millionaires' fortunes. Hopefully, our in-depth analysis demonstrates that it is very easy to extend the present example either by increasing the capacity $m$ of the registers or by increasing the number of millionaires $n$.

\section{Efficiency and security analysis} \label{sec: Efficiency & Security Analysis}

The current section is devoted to an extensive efficiency and security analysis of the GHZ$_{ 3 }$MQPEC protocol. We first consider its efficiency, which, we believe, is one of the strong points of the protocol.

\subsection{Efficiency} \label{subsec: Efficiency}

A typical measure of the qubit efficiency of quantum protocols is the ratio $\eta$ of the total number of transmitted ``useful'' classical bits to the total number of utilized qubits \cite{Tsai2011, Hwang2011}. For multiparty quantum private comparison protocols this measure must be suitably adjusted and defined as follows (see for instance \cite{Huang2023, Hou2024})
\begin{align} \label{eq: GHZ$_{ 3 }$MQPEC Eta Efficiency Definition}
	\eta
	=
	\frac { \eta_{ cb } } { \eta_{ tq } }
	\ ,
\end{align}
where $\eta_{ cb }$ is the total number of (classical) bits that are necessary to represent the fortunes the millionaires want to compare for equality, and $\eta_{ tq }$ is the total number of qubits utilized by the protocol. At this point, we must clarify that the measure $\eta_{ tq }$ doesn't include the decoys, as is the norm in the literature.

In the case of the GHZ$_{ 3 }$MQPEC protocol, we have assumed that each millionaire requires $m$ bits to store her fortune. Taking into account the fact that there are $n$ millionaires, we see that
\begin{align} \label{eq: GHZ$_{ 3 }$MQPEC Eta_cb Efficiency}
	\eta_{ cb }
	=
	n m
	\ .
\end{align}
To accomplish this task our protocol requires $3 m + 2$ qubits for each millionaire, which means that for $n$ millionaires it requires
\begin{align} \label{eq: GHZ$_{ 3 }$MQPEC Eta_tq Efficiency}
	\eta_{ tq }
	=
	3 n m + 2 n
\end{align}
qubits in total. Therefore, the efficiency $\eta$ of the GHZ$_{ 3 }$MQPEC protocol is
\begin{align} \label{eq: GHZ$_{ 3 }$MQPEC Eta Efficiency}
	\eta
	=
	\frac { n m } { 3 n m + 2 n }
	=
	\frac { m } { 3 m + 2 }
	\approx
	\frac { m } { 3 m }
	=
	\frac { 1 } { 3 }
	=
	33.33\%
	\ .
\end{align}
In the derivation of the above result we have used the fact that for large values of $m$, a realistic assumption in a real-life scenario, it holds that $3 m + 2 \approx 3 m$. These efficiency results are what is to be expected in multiparty quantum private comparison protocols that utilize GHZ states. For an extensive and detailed comparative analysis of the efficiency of many similar protocols, where this fact can be corroborated, we refer to \cite{Banerjee2012, Joy2017, Song2019, Huang2023, Hou2024}. In closing this subsection, we point out that the quantum resources required by our GHZ$_{ 3 }$MQPEC protocol, as given by \eqref{eq: GHZ$_{ 3 }$MQPEC Eta_tq Efficiency} are linear both in the number of millionaires $n$, and the number of qubits $m$ that suffice to represent in binary any of the millionaires' fortunes, as expressed below
\begin{align} \label{eq: GHZ$_{ 3 }$MQPEC Linearity of Quantum Qubits}
	\eta_{ tq }
	=
	\Theta ( n + m )
	\ .
\end{align}
This proves the ability of the GHZ$_{ 3 }$MQPEC protocol to scale in the most efficient way possible with both parameters $n$ and $m$, and, thus, be particularly amenable to practical implementation.

\subsection{Security} \label{subsec: Security}

In this subsection we focus on security issues. We follow the general line of thought advocated in recent works addressing security issues of quantum protocols, such as \cite{Wolf2021} and \cite{Renner2023}, with the necessary adjustments for our particular setting. Consequently, it is expedient to distinguish between \textit{external} and \textit{participants'} attacks. By external attack we refer to an attack by an outside eavesdropper, namely Eve, whereas by participants' attack we mean an attack by any of the $n + 2$ players that participate in the protocol.

\subsubsection{External attack} \label{subsubsec: External Attack}

Before we lay out the formal analysis regarding the possible attacks that Eve may use in order to compromise the GHZ$_{ 3 }$MQPEC protocol and gain secret information, it is essential to recall two standard detection techniques that are among the first line of defense against malicious adversaries.

\begin{enumerate}
	\renewcommand\labelenumi{(\textbf{DT}$_\theenumi$)}
	\item	\textbf{Eavesdropping detection.} To detect the presence of a possible eavesdropper, the source that produces and distributes the entangled particles, the $\ket{ GHZ_{ 3 } }$ triplets in our case, also produces decoy photons randomly chosen from one of the $\ket{ 0 }, \ket{ 1 }, \ket{ + }, \ket{ - }$ states. These photons are randomly inserted into the transmission sequences. Eve, by meddling with the transmission sequences, and unaware of the positions of the decoys, will, invariably, cause errors that are easily detectable by the legitimate players upon a typical verification check. This technique, which is referred to as the \textit{decoy technique}, has been extensively studied in the literature \cite{Deng2008, Yang2009, Tseng2011, Chang2012, Hung2016, Ye2018, Wu2021, Hou2024}.
	\item	\textbf{Entanglement validation.} The successful implementation of any entanglement-based protocol hinges upon the existence of entanglement. Without guaranteed entanglement, the protocol's functionality is compromised. The entanglement validation procedure can result into two distinct outcomes. If entanglement is successfully ascertained, the protocol can proceed to confidently accomplish its intended task. Failure to validate entanglement indicates the absence of the necessary entanglement. This could stem from either noisy quantum channels or malicious interference from an adversary. Regardless of the cause, the only viable solution is to halt the ongoing execution and commence the entire procedure anew, after implementing corrective measures. Given its significance, entanglement validation has been thoroughly studied in the literature. Our protocol adheres to the sophisticated methodologies outlined in prior works, including \cite{Neigovzen2008, Feng2019, Wang2022a, Yang2022, Qu2023, Ikeda2023c}.
\end{enumerate}

Let us now consider the most well-known attacks that the outside eavesdropper Eve may employ.

\begin{enumerate}
	\renewcommand\labelenumi{(\textbf{Attack}$_\theenumi$)}
	\item	\textbf{Measure and Resend}. In this type of attack, Eve intercepts the $\ket{ GHZ_{ 3 } }$ qubits during their transmission from Trent to Sophia and Alice$_{ i }$, $0 \leq i \leq n - 1$, measures them and then sends them back to their intended recipients. By doing so, Eve will fail to discover any information because at this phase the $\ket{ GHZ_{ 3 } }$ triplets do not be carry any information. Moreover, by resorting to the decoy technique, the legitimate players will easily detect a malicious presence. Consequently, the protocol is completely impervious to such an attack.
	\item	\textbf{Intercept and Resend}. In this case, Eve's strategy is to intercept the $\ket{ GHZ_{ 3 } }$ particles during their transmission from Trent to Sophia and Alice$_{ i }$, $0 \leq i \leq n - 1$. Eve is not able to clone them because of the no-cloning theorem. Accordingly, she prepares and transmits new qubits to the intended recipient. As, we have pointed out above, during the transmission phase of the protocol, the $\ket{ GHZ_{ 3 } }$ triplets carry no information whatsoever, so, by measuring them, Eve doesn't gain any information. Furthermore, the decoy technique, allows the legitimate players to infer the presence of an eavesdropper. So, the GHZ$_{ 3 }$MQPEC protocol is also safe from such an attack.
	\item	\textbf{Entangle and Measure}. In this type of attack, Eve begins by intercepting the qubits of the $\ket{ GHZ_{3} }$ triplets during their transmission. In contrast to the previous types of attack, in this case Eve does not measure them, but entangles them with her ancilla state, and then sends the corresponding qubits to Sophia and Alice$_{ i }$. Then, Eve waits until the protocol completes before measuring her qubits, hoping to gain useful information. Unfortunately for Eve, the outcome of her actions is, instead of having $\ket{ GHZ_{ 3 } }$ triplets evenly distributed among Trent, Sophia and Alice$_{ i }$, to end up with $m$ $\ket{ GHZ_{ 4 } }$ quadruples evenly distributed among Trent, Sophia, Alice$_{ i }$ and Eve. Therefore, during the classical part of the protocol, when Sophia and Alice$_{ i }$ send their measurements to Trent through the public channel, Eve will fail to compute the correct modulo $2$ sum because in order to achieve this, she needs Trent's private bit vector $\mathbf{ y }_{ 2 }$ that Trent never transmits through the public channel. Therefore, in this case too, Eve will fail to gain any information.
	\item	\textbf{PNS}. The photon number splitting attack (PNS), first introduced in \cite{huttner1995quantum} and later analyzed in \cite{lutkenhaus2000security, brassard2000limitations}, is currently regarded as one of the most effective attack strategies that Eve can employ against any quantum protocol. Contemporary photon sources exhibit a known flaw: occasionally they may produce multiple identical photons instead of just one. So, Eve may intercept pulses emanating from Trent for the distribution of the $\ket{ GHZ_{ 3 } }$ triplets, keep one photon from the multi-photon pulse for herself and send the remaining photons to Sophia and Alice$_{ i }$ without being detected during the transmission phase. Nonetheless, this situation resembles the Entangle and Measure attack analyzed above. Again, instead of $\ket{ GHZ_{ 3 } }$ triplets evenly distributed among the three legitimate actors, there are $\ket{ GHZ_{ 4 } }$ quadruples evenly distributed among Trent, Sophia, Alice$_{ i }$ and Eve. Eve becomes an active fourth player in the game, but fails to uncover any useful information for the reason explained above.
\end{enumerate}

\subsubsection{Participants' attack} \label{subsubsec: Participants' Attack}

We now consider the case where one of the players of this game tries to gain some information about the fortune of one of the millionaires. It is convenient to distinguish the following cases depending on the dishonest participant that attempts to gain unauthorized information.

\begin{itemize}
	\item[\textbf{Alice}$_{ i }$] Let us assume that Alice$_{ i }$, $0 \leq i \leq n - 1$, intercepts the $\ket{ GHZ_{ 3 } }$ qubits during their transmission from Trent to Sophia and another Alice$_{ j }$, $0 \leq i \neq j \leq n - 1$. Alice$_{ i }$ has the option to measure them and send them back to their intended recipients, or transmit new, previously prepared, qubits to their intended recipients. In both cases, Alice$_{ i }$ will not discover any information because at this phase the $\ket{ GHZ_{ 3 } }$ triplets do not be carry any information. Moreover, the use of the decoy technique will allow the legitimate players to infer that someone has tampered with the transmitted sequences.
	
	Another possibility is Alice$_{ i }$ to intercept the qubits from the $\ket{ GHZ_{3} }$ triplets during their transmission from Trent to Sophia and Alice$_{ j }$, $i \neq j$, and, without measuring them, entangle them with her ancilla state. Alice$_{ i }$ sends corresponding qubits to Sophia and Alice$_{ j }$, and waits until the protocol completes before measuring her qubits, in the hope of gaining useful information. The critical flaw in Alice$_{ i }$'s strategy is that her actions result in having $\ket{ GHZ_{ 4 } }$ quadruples evenly distributed among Trent, Sophia, Alice$_{ j }$ and Alice$_{ i }$, instead of having $\ket{ GHZ_{ 3 } }$ triplets evenly distributed among Trent, Sophia and Alice$_{ j }$. Obviously, during the classical part of the protocol, when Sophia and Alice$_{ j }$ send their measurements to Trent through the public channel, Alice$_{ i }$ will fail to compute the correct modulo $2$ sum because in order to achieve this, she needs Trent's private bit vector $\mathbf{ y }_{ 2 }$. Recall that Trent never divulges $\mathbf{ y }_{ 2 }$ or the modulo $2$ sum $\mathbf{ y }_{ 2 } \oplus \mathbf{ y }_{ 1 } \oplus \mathbf{ y }_{ 0 }$ to any other in-game player or outside eavesdropper, as ordained by (\textbf{A}$_{ 4 }$) and (\textbf{D}$_{ 3 }$). Once again, Alice$_{ i }$ fails to gain any information. Therefore, the GHZ$_{ 3 }$MQPEC protocol is immune to attacks by Alice$_{ i }$, $0 \leq i \leq n - 1$.
	\item[\textbf{Sophia}]  Let us assume that Sophia intercepts the $\ket{ GHZ_{ 3 } }$ particles during their transmission from Trent to Alice$_{ i }$. Sophia may measure them and send them back to Alice$_{ i }$, or transmit to Alice$_{ i }$ new previously prepared qubits. In any of the previous cases, Sophia will not discover any information because at this phase the $\ket{ GHZ_{ 3 } }$ triplets do not be carry any information. Moreover, the use of the decoy technique will allow the other legitimate players to detect the tampering.
	
	Another possibility is Sophia to intercept the qubits from the $\ket{ GHZ_{3} }$ triplets during their transmission from Trent to Alice$_{ i }$ and entangle them with her ancilla state, without measuring them. Sophia sends corresponding qubits to Alice$_{ i }$ and waits until the protocol has completed before measuring her qubits, hoping to gain some information. The problem with this strategy is that the result of Sophia's actions leads to having $\ket{ GHZ_{ 4 } }$ quadruples evenly distributed among Trent, Sophia and Alice$_{ i }$, instead of having $\ket{ GHZ_{ 3 } }$ triplets. Obviously, during the classical part of the protocol, when Sophia and Alice$_{ i }$ send their measurements to Trent through the public channel, Sophia will fail to compute the correct modulo $2$ sum because in order to achieve this, she needs Trent's private bit vector $\mathbf{ y }_{ 2 }$. But Trent never divulges $\mathbf{ y }_{ 2 }$ or the modulo $2$ sum $\mathbf{ y }_{ 2 } \oplus \mathbf{ y }_{ 1 } \oplus \mathbf{ y }_{ 0 }$ to any other in-game player or outside eavesdropper, as dictated by (\textbf{A}$_{ 4 }$) and (\textbf{D}$_{ 3 }$). Consequently, the GHZ$_{ 3 }$MQPEC protocol is impervious to attacks by Sophia.
	\item[\textbf{Trent}] Trent can't surmise Sophia's secret number $\mathbf{ s }$ because Sophia, as stipulated in (\textbf{A}$_{ 3 }$), never reveals her secret number $\mathbf{ s }$. Moreover, the fundamental premise of the protocol is that Alice$_{ i }$, $0 \leq i \leq n - 1$, never divulges her private fortune $\mathbf{ f }_{ i }$. Hence, although Trent knows the modulo $2$ sum $\mathbf{ s } \oplus \mathbf{ f }_{ i }$, he can infer neither $\mathbf{ s }$ nor $\mathbf{ f }_{ i }$. Thus, the GHZ$_{ 3 }$MQPEC protocol cannot be compromised by Trent.
\end{itemize}

The above succinct security analysis demonstrates that the GHZ$_{ 3 }$MQPEC protocol is information-theoretically secure.

\section{Discussion and conclusions} \label{sec: Discussion and Conclusions}

This paper introduces the innovative entanglement-based GHZ$_{ 3 }$MQPEC protocol that accomplishes multiparty quantum private comparison leveraging maximally entangled $\ket{ GHZ_{ 3 } }$ triplets. We list below some of the most important advantages of the protocol.

\begin{itemize}
	[ left = 0.00 cm, labelsep = 0.50 cm ]
	\item	The primary aim of this protocol is to be easily executable by modern quantum computers. This is achieved by employing only $\ket{ GHZ_{ 3 } }$ triplets, regardless of the number of participants. While more complex multi-particle entangled states and high-dimensional quantum states are possible, they are difficult to produce with current quantum technology, resulting in longer preparation times and increased complexity, especially in scenarios with many participants. By exclusively using $\ket{ GHZ_{ 3 } }$ states, which are simpler to produce than Bell states, these issues are mitigated, advancing the practical implementation of the protocol.
	\item	A key quantitative feature of this protocol is that the required quantum resources scale linearly with both the number of participants and the amount of information being compared, further demonstrating its practicality. Notably, the GHZ$_{ 3 }$MQPEC protocol can be executed in both parallel and sequential modes. While the quantum operations are intended to run in parallel, sequential execution is also possible if quantum resources are insufficient, allowing the participants to be divided into smaller groups for sequential processing.
	\item	Additionally, the protocol does not depend on a quantum signature scheme or pre-shared keys, which simplifies and reduces its cost. Implementation is further streamlined as all participants use identical private quantum circuits consisting of Hadamard and CNOT gates, making it straightforward to implement on contemporary quantum computers.
	\item	Crucially, the protocol ensures that external parties cannot access any information about the participants' values, and participants themselves cannot learn each other’s secret numbers. The protocol is demonstrated to be information-theoretically secure.
\end{itemize}

In closing, let us consider the special case where $n = 2$, that is when there are only $2$ millionaires, say Alice and Bob. As we have mentioned in section \ref{sec: The Setup}, in such a case Sophia becomes unnecessary and Trent alone suffices. The GHZ$_{ 3 }$MQPEC protocol can be easily simplified to function without the presence of Sophia, using the quantum circuit shown in Figure \ref{fig: The Special Quantum Circuit of the GHZ$_{ 3 }$MQPEC Protocol for $2$ Millionaires}. Obviously, the whole mathematical description of the protocol, its quantitative characteristics, and its security remain unaffected.

\begin{tcolorbox}
	[
		enhanced,
		breakable,
		grow to left by = 0.50 cm,
		grow to right by = 0.00 cm,
		colback = WordVeryLightTeal!12,			
		enhanced jigsaw,						
		sharp corners,
		toprule = 1.0 pt,
		bottomrule = 1.0 pt,
		leftrule = 0.1 pt,
		rightrule = 0.1 pt,
		sharp corners,
		center title,
		fonttitle = \bfseries
	]
	\begin{figure}[H]
		\centering
		\begin{tikzpicture}[ scale = 0.900, transform shape ]
			\begin{yquant}
				nobit AUX_B_0;
				[ name = Bob ] qubits { $BIR$ } BIR;
				qubit { $BOR$: \ $\ket{ - }$ } BOR;
				nobit AUX_B_1;
				[ name = spaceBA, register/minimum height = 8 mm ] nobit spaceBA;
				nobit AUX_A_0;
				[ name = Alice ] qubits { $AIR$ } AIR;
				qubit { $BOR$: \ $\ket{ - }$ } AOR;
				nobit AUX_A_1;
				nobit AUX_A_2;
				[ name = spaceAT, register/minimum height = 8 mm ] nobit spaceAT;
				nobit AUX_T_0;
				[ name = Trent ] qubits { $TIR$ } TIR;
				nobit AUX_T_1;
				nobit AUX_T_2;
				[ name = Ph0, WordBlueDarker, line width = 0.50 mm, label = { [ label distance = 0.20 cm ] north: Initial State } ]
				barrier ( - ) ;
				[ draw = WordGoldAccent1Lighter40, fill = MagentaLight!, radius = 0.7 cm ] box {\color{white} \Large \sf{U}$_{ \mathbf{ f }_{ B } }$} (BIR - BOR);
				[ draw = WordGoldAccent1Lighter40, fill = MagentaLight, radius = 0.7 cm ] box {\color{white} \Large \sf{U}$_{ \mathbf{ f }_{ A } }$} (AIR - AOR);
				[ name = Ph1, WordBlueDarker, line width = 0.50 mm, label = { [ label distance = 0.20 cm ] north: Phase 1 } ]
				barrier ( - ) ;
				[ draw = gray, fill = gray, radius = 0.6 cm ] box {\color{white} \Large \sf{H}$^{ \otimes n }$} BIR;
				[ draw = gray, fill = gray, radius = 0.6 cm ] box {\color{white} \Large \sf{H}$^{ \otimes n }$} AIR;
				[ draw = gray, fill = gray, radius = 0.6 cm ] box {\color{white} \Large \sf{H}$^{ \otimes n }$} TIR;
				[ name = Ph2, WordBlueDarker, line width = 0.50 mm, label = { [ label distance = 0.20 cm ] north: Phase 2 } ]
				barrier ( - ) ;
				[ line width = .250 mm, draw = white, fill = gray, radius = 0.6 cm ] measure BIR;
				[ line width = .250 mm, draw = white, fill = gray, radius = 0.6 cm ] measure AIR;
				[ line width = .250 mm, draw = white, fill = gray, radius = 0.6 cm ] measure TIR;
				[ name = Ph3, WordBlueDarker, line width = 0.50 mm, label = { [ label distance = 0.20 cm ] north: Measurement } ]
				barrier ( - ) ;
				output { $\ket{ \mathbf{ y }_{ 0 } }$ } BIR;
				output { $\ket{ \mathbf{ y }_{ 1 } }$ } AIR;
				output { $\ket{ \mathbf{ y }_{ 2 } }$ } TIR;
				\node [ below = 5.00 cm ] at (Ph0) { $\ket{ \psi_{ 0 } }$ };
				\node [ below = 5.00 cm ] at (Ph1) { $\ket{ \psi_{ 1 } }$ };
				\node [ below = 5.00 cm ] at (Ph2) { $\ket{ \psi_{ 2 } }$ };
				\node [ below = 5.00 cm ] at (Ph3) { $\ket{ \psi_{ f } }$ };
				\node
				[
				charlie,
				scale = 1.50,
				anchor = center,
				left = 0.50 cm of Bob,
				label = { [ label distance = 0.00 cm ] west: Bob }
				]
				() { };
				\node
				[
				alice,
				scale = 1.50,
				anchor = center,
				left = 0.50 cm of Sophia,
				label = { [ label distance = 0.00 cm ] west: Alice }
				]
				() { };
				\node
				[
				maninblack,
				scale = 1.50,
				anchor = center,
				left = 0.50 cm of Trent,
				label = { [ label distance = 0.00 cm ] west: Trent }
				]
				() { };
				\begin{scope} [ on background layer ]
					\node [ above right = - 0.30 cm and 0.70 cm of spaceBA, rectangle, fill = WordAquaLighter60, text width = 10.00 cm, align = center, minimum height = 10 mm ] { \bf Possibly Spatially Separated };
					\node [ above right = - 0.30 cm and 0.70 cm of spaceAT, rectangle, fill = WordAquaLighter60, text width = 10.00 cm, align = center, minimum height = 10 mm ] { \bf Possibly Spatially Separated };
				\end{scope}
			\end{yquant}
			\node
			[
			above right = 3.00 cm and 5.750 cm of Bob,
			anchor = center,
			shade,
			top color = GreenTeal, bottom color = black,
			rectangle,
			text width = 11.00 cm,
			align = center
			]
			(Label)
			{ \color{white}
				The quantum circuit for the special case of $2$ millionaires
			};
			\node [ anchor = center, below = 1.00 cm of Trent ] (PhantomNode) { };
			\scoped [ on background layer ]
			\draw
			[ RedPurple, -, >=stealth, line width = 0.75 mm, decoration = coil, decorate ]
			( $ (Trent.east) + ( 0.5 mm, 0 mm ) $ ) node [ circle, fill, minimum size = 1.5 mm ] () {} -- ( $ (Alice.east) + ( 0.5 mm, 0 mm ) $ ) node [ circle, fill, minimum size = 1.5 mm ] () {} -- ( $ (Bob.east) + ( 0.5 mm, 0 mm ) $ ) node [ circle, fill, minimum size = 1.5 mm ] () {};
		\end{tikzpicture}
		\caption{The above circuit embeds Alice and Bob's fortune $\mathbf{ f }_{ A }$ and $\mathbf{ f }_{ B }$ into the global state of the system. The private circuits operated by Alice, Bob, and Trent, are linked due to entanglement, indicated by the wavy red line connecting $TIR, AIR$ and $BIR$, and form one composite system.}
		\label{fig: The Special Quantum Circuit of the GHZ$_{ 3 }$MQPEC Protocol for $2$ Millionaires}
	\end{figure}
\end{tcolorbox}
\bibliographystyle{ieeetr}
\bibliography{GHZ3MQPEC}

\begin{thebibliography}{10}

\bibitem{IBMEagle}
J.~Chow, O.~Dial, and J.~Gambetta, ``{IBM} {Quantum} breaks the 100-qubit
  processor barrier.''
  \url{https://www.ibm.com/quantum/blog/127-qubit-quantum-processor-eagle},
  2021.
\newblock Accessed: 2024-03-02.

\bibitem{IBMOsprey}
I.~Newsroom, ``{IBM} unveils 400 qubit-plus quantum processor.''
  \url{https://newsroom.ibm.com/2022-11-09-IBM-Unveils-400-Qubit-Plus-Quantum-Processor-and-Next-Generation-IBM-Quantum-System-Two},
  2022.
\newblock Accessed: 2024-03-02.

\bibitem{IBMCondor}
J.~Gambetta, ``The hardware and software for the era of quantum utility is
  here.'' \url{https://www.ibm.com/quantum/blog/quantum-roadmap-2033}, 2023.
\newblock Accessed: 2024-03-02.

\bibitem{Bennett1984}
C.~H. Bennett and G.~Brassard, ``Quantum cryptography: Public key distribution
  and coin tossing,'' in {\em Proceedings of the IEEE International Conference
  on Computers, Systems, and Signal Processing}, pp.~175--179, {IEEE} Computer
  Society Press, 1984.

\bibitem{Ekert1991}
A.~K. Ekert, ``Quantum cryptography based on bell's theorem,'' {\em Physical
  Review Letters}, vol.~67, no.~6, pp.~661--663, 1991.

\bibitem{Bennett1992}
C.~H. Bennett, G.~Brassard, and N.~D. Mermin, ``Quantum cryptography without
  bell's theorem,'' {\em Physical Review Letters}, vol.~68, no.~5,
  pp.~557--559, 1992.

\bibitem{Bennett2014}
C.~H. Bennett and G.~Brassard, ``Quantum cryptography: Public key distribution
  and coin tossing,'' {\em Theoretical Computer Science}, vol.~560, pp.~7--11,
  2014.

\bibitem{Ampatzis2021}
M.~Ampatzis and T.~Andronikos, ``{QKD} based on symmetric entangled
  bernstein-vazirani,'' {\em Entropy}, vol.~23, no.~7, p.~870, 2021.

\bibitem{Hillery1999}
M.~Hillery, V.~Bu{\v{z}}ek, and A.~Berthiaume, ``Quantum secret sharing,'' {\em
  Physical Review A}, vol.~59, no.~3, p.~1829, 1999.

\bibitem{Cleve1999}
R.~Cleve, D.~Gottesman, and H.-K. Lo, ``How to share a quantum secret,'' {\em
  Physical Review Letters}, vol.~83, no.~3, p.~648, 1999.

\bibitem{Karlsson1999}
A.~Karlsson, M.~Koashi, and N.~Imoto, ``Quantum entanglement for secret sharing
  and secret splitting,'' {\em Physical Review A}, vol.~59, no.~1, p.~162,
  1999.

\bibitem{Ampatzis2022}
M.~Ampatzis and T.~Andronikos, ``A symmetric extensible protocol for quantum
  secret sharing,'' {\em Symmetry}, vol.~14, no.~8, p.~1692, 2022.

\bibitem{Bennett1993}
C.~H. Bennett, G.~Brassard, C.~Cr{\'{e}}peau, R.~Jozsa, A.~Peres, and W.~K.
  Wootters, ``Teleporting an unknown quantum state via dual classical and
  einstein-podolsky-rosen channels,'' {\em Physical Review Letters}, vol.~70,
  no.~13, pp.~1895--1899, 1993.

\bibitem{Bouwmeester1997}
D.~Bouwmeester, J.-W. Pan, K.~Mattle, M.~Eibl, H.~Weinfurter, and A.~Zeilinger,
  ``Experimental quantum teleportation,'' {\em Nature}, vol.~390, no.~6660,
  pp.~575--579, 1997.

\bibitem{Deng2003}
F.-G. Deng, G.~L. Long, and X.-S. Liu, ``Two-step quantum direct communication
  protocol using the einstein-podolsky-rosen pair block,'' {\em Physical Review
  A}, vol.~68, no.~4, p.~042317, 2003.

\bibitem{Deng2004}
F.-G. Deng and G.~L. Long, ``Secure direct communication with a quantum
  one-time pad,'' {\em Physical Review A}, vol.~69, no.~5, p.~052319, 2004.

\bibitem{Wang2005}
C.~Wang, F.-G. Deng, Y.-S. Li, X.-S. Liu, and G.~L. Long, ``Quantum secure
  direct communication with high-dimension quantum superdense coding,'' {\em
  Physical Review A}, vol.~71, no.~4, p.~044305, 2005.

\bibitem{Yang2009}
Y.-G. Yang and Q.-Y. Wen, ``An efficient two-party quantum private comparison
  protocol with decoy photons and two-photon entanglement,'' {\em Journal of
  Physics A: Mathematical and Theoretical}, vol.~42, no.~5, p.~055305, 2009.

\bibitem{Yao1982}
A.~C. Yao, ``Protocols for secure computations,'' in {\em 23rd Annual Symposium
  on Foundations of Computer Science (sfcs 1982)}, IEEE, 1982.

\bibitem{Yao1986}
A.~C.-C. Yao, ``How to generate and exchange secrets,'' in {\em 27th Annual
  Symposium on Foundations of Computer Science (sfcs 1986)}, IEEE, 1986.

\bibitem{Boudot2001}
F.~Boudot, B.~Schoenmakers, and J.~Traoré, ``A fair and efficient solution to
  the socialist millionaires’ problem,'' {\em Discrete Applied Mathematics},
  vol.~111, no.~1–2, pp.~23--36, 2001.

\bibitem{Chen2010}
X.-B. Chen, G.~Xu, X.-X. Niu, Q.-Y. Wen, and Y.-X. Yang, ``An efficient
  protocol for the private comparison of equal information based on the triplet
  entangled state and single-particle measurement,'' {\em Optics
  Communications}, vol.~283, no.~7, pp.~1561--1565, 2010.

\bibitem{Liu2011}
W.~Liu, Y.-B. Wang, and Z.-T. Jiang, ``An efficient protocol for the quantum
  private comparison of equality with w state,'' {\em Optics Communications},
  vol.~284, no.~12, pp.~3160--3163, 2011.

\bibitem{Tseng2011}
H.-Y. Tseng, J.~Lin, and T.~Hwang, ``New quantum private comparison protocol
  using epr pairs,'' {\em Quantum Information Processing}, vol.~11, no.~2,
  pp.~373--384, 2011.

\bibitem{Liu2011a}
W.~Liu, Y.-B. Wang, Z.-T. Jiang, and Y.-Z. Cao, ``A protocol for the quantum
  private comparison of equality with $\chi$-type state,'' {\em International
  Journal of Theoretical Physics}, vol.~51, no.~1, pp.~69--77, 2011.

\bibitem{Jia2011a}
H.-Y. Jia, Q.-Y. Wen, Y.-B. Li, and F.~Gao, ``Quantum private comparison using
  genuine four-particle entangled states,'' {\em International Journal of
  Theoretical Physics}, vol.~51, no.~4, pp.~1187--1194, 2011.

\bibitem{Liu2012}
W.~Liu, Y.-B. Wang, Z.-T. Jiang, Y.-Z. Cao, and W.~Cui, ``New quantum private
  comparison protocol using $\chi$-type state,'' {\em International Journal of
  Theoretical Physics}, vol.~51, no.~6, pp.~1953--1960, 2012.

\bibitem{Liu2012a}
W.~Liu and Y.-B. Wang, ``Quantum private comparison based on ghz entangled
  states,'' {\em International Journal of Theoretical Physics}, vol.~51,
  no.~11, pp.~3596--3604, 2012.

\bibitem{Ji2016}
Z.-X. Ji and T.-Y. Ye, ``Quantum private comparison of equal information based
  on highly entangled six-qubit genuine state,'' {\em Communications in
  Theoretical Physics}, vol.~65, no.~6, pp.~711--715, 2016.

\bibitem{Chou2016}
W.-H. Chou, T.~Hwang, and J.~Gu, ``Semi-quantum private comparison protocol
  under an almost-dishonest third party,'' 2016.

\bibitem{He2023}
Z.~He and X.~Lou, ``Security analysis and improvement in a semi-quantum private
  comparison without pre-shared key,'' {\em Quantum Information Processing},
  vol.~22, no.~3, 2023.

\bibitem{Chen2012}
X.-B. Chen, Y.~Su, X.-X. Niu, and Y.-X. Yang, ``Efficient and feasible quantum
  private comparison of equality against the collective amplitude damping
  noise,'' {\em Quantum Information Processing}, vol.~13, no.~1, pp.~101--112,
  2012.

\bibitem{Zi2013}
W.~Zi, F.~Guo, Y.~Luo, S.~Cao, and Q.~Wen, ``Quantum private comparison
  protocol with the random rotation,'' {\em International Journal of
  Theoretical Physics}, vol.~52, no.~9, pp.~3212--3219, 2013.

\bibitem{Ye2017}
T.-Y. Ye, ``Quantum private comparison via cavity qed,'' {\em Communications in
  Theoretical Physics}, vol.~67, no.~2, p.~147, 2017.

\bibitem{Hou2024}
M.~Hou and Y.~Wu, ``Single-photon-based quantum secure protocol for the
  socialist millionaires’ problem,'' {\em Frontiers in Physics}, vol.~12,
  2024.

\bibitem{Liu2013}
W.~Liu, C.~Liu, H.~Wang, and T.~Jia, ``Quantum private comparison: A review,''
  {\em IETE Technical Review}, vol.~30, no.~5, p.~439, 2013.

\bibitem{Jia2011}
H.-Y. Jia, Q.-Y. Wen, T.-T. Song, and F.~Gao, ``Quantum protocol for
  millionaire problem,'' {\em Optics Communications}, vol.~284, no.~1,
  pp.~545--549, 2011.

\bibitem{Lin2012}
S.~Lin, Y.~Sun, X.-F. Liu, and Z.-Q. Yao, ``Quantum private comparison protocol
  with d-dimensional bell states,'' {\em Quantum Information Processing},
  vol.~12, no.~1, pp.~559--568, 2012.

\bibitem{Zhang2013}
W.-W. Zhang, D.~Li, K.-J. Zhang, and H.-J. Zuo, ``A quantum protocol for
  millionaire problem with bell states,'' {\em Quantum Information Processing},
  vol.~12, no.~6, pp.~2241--2249, 2013.

\bibitem{Ye2018}
C.-Q. Ye and T.-Y. Ye, ``Multi-party quantum private comparison of size
  relation with d-level single-particle states,'' {\em Quantum Information
  Processing}, vol.~17, no.~10, 2018.

\bibitem{Cao2019}
H.~Cao, W.~Ma, L.~Lü, Y.~He, and G.~Liu, ``Multi-party quantum privacy
  comparison of size based on d-level ghz states,'' {\em Quantum Information
  Processing}, vol.~18, no.~9, 2019.

\bibitem{Wu2021}
W.~Wu and Y.~Zhao, ``Quantum private comparison of size using d-level bell
  states with a semi-honest third party,'' {\em Quantum Information
  Processing}, vol.~20, no.~4, 2021.

\bibitem{Chang2012}
Y.-J. Chang, C.-W. Tsai, and T.~Hwang, ``Multi-user private comparison protocol
  using ghz class states,'' {\em Quantum Information Processing}, vol.~12,
  no.~2, pp.~1077--1088, 2001.

\bibitem{Liu2013a}
W.~Liu, Y.-B. Wang, and X.-M. Wang, ``Multi-party quantum private comparison
  protocol using d-dimensional basis states without entanglement swapping,''
  {\em International Journal of Theoretical Physics}, vol.~53, no.~4,
  pp.~1085--1091, 2013.

\bibitem{Huang2015}
S.-L. Huang, T.~Hwang, and P.~Gope, ``Multi-party quantum private comparison
  with an almost-dishonest third party,'' {\em Quantum Information Processing},
  vol.~14, no.~11, pp.~4225--4235, 2015.

\bibitem{Hung2016}
S.-M. Hung, S.-L. Hwang, T.~Hwang, and S.-H. Kao, ``Multiparty quantum private
  comparison with almost dishonest third parties for strangers,'' {\em Quantum
  Information Processing}, vol.~16, no.~2, 2016.

\bibitem{Zhang2023}
J.-W. Zhang, G.~Xu, X.-B. Chen, Y.~Chang, and Z.-C. Dong, ``Improved multiparty
  quantum private comparison based on quantum homomorphic encryption,'' {\em
  Physica A: Statistical Mechanics and its Applications}, vol.~610, p.~128397,
  2023.

\bibitem{Colbeck2007}
R.~Colbeck, ``Impossibility of secure two-party classical computation,'' {\em
  Physical Review A}, vol.~76, no.~6, p.~062308, 2007.

\bibitem{Crepeau2002}
C.~Crépeau, D.~Gottesman, and A.~Smith, ``Secure multi-party quantum
  computation,'' in {\em Proceedings of the thiry-fourth annual ACM symposium
  on Theory of computing}, STOC02, ACM, 2002.

\bibitem{Lo1997}
H.-K. Lo, ``Insecurity of quantum secure computations,'' {\em Physical Review
  A}, vol.~56, no.~2, pp.~1154--1162, 1997.

\bibitem{Meyer1999}
D.~A. Meyer, ``Quantum strategies,'' {\em Physical Review Letters}, vol.~82,
  no.~5, p.~1052, 1999.

\bibitem{Eisert1999}
J.~Eisert, M.~Wilkens, and M.~Lewenstein, ``Quantum games and quantum
  strategies,'' {\em Physical Review Letters}, vol.~83, no.~15, p.~3077, 1999.

\bibitem{Andronikos2018}
T.~Andronikos, A.~Sirokofskich, K.~Kastampolidou, M.~Varvouzou, K.~Giannakis,
  and A.~Singh, ``Finite automata capturing winning sequences for all possible
  variants of the {PQ} penny flip game,'' {\em Mathematics}, vol.~6, p.~20, Feb
  2018.

\bibitem{Andronikos2021}
T.~Andronikos and A.~Sirokofskich, ``The connection between the {PQ} penny flip
  game and the dihedral groups,'' {\em Mathematics}, vol.~9, no.~10, p.~1115,
  2021.

\bibitem{Andronikos2022a}
T.~Andronikos, ``Conditions that enable a player to surely win in sequential
  quantum games,'' {\em Quantum Information Processing}, vol.~21, no.~7, 2022.

\bibitem{Kastampolidou2023a}
K.~Kastampolidou and T.~Andronikos, ``Quantum tapsilou—a quantum game
  inspired by the traditional greek coin tossing game tapsilou,'' {\em Games},
  vol.~14, no.~6, p.~72, 2023.

\bibitem{Giannakis2019}
K.~Giannakis, G.~Theocharopoulou, C.~Papalitsas, S.~Fanarioti, and
  T.~Andronikos, ``Quantum conditional strategies and automata for prisoners'
  dilemmata under the {EWL} scheme,'' {\em Applied Sciences}, vol.~9, p.~2635,
  Jun 2019.

\bibitem{Andronikos2022}
T.~Andronikos and M.~Stefanidakis, ``A two-party quantum parliament,'' {\em
  Algorithms}, vol.~15, no.~2, p.~62, 2022.

\bibitem{Ampatzis2023}
M.~Ampatzis and T.~Andronikos, ``Quantum secret aggregation utilizing a network
  of agents,'' {\em Cryptography}, vol.~7, no.~1, p.~5, 2023.

\bibitem{Andronikos2023}
T.~Andronikos and A.~Sirokofskich, ``An entanglement-based protocol for
  simultaneous reciprocal information exchange between 2 players,'' {\em
  Electronics}, vol.~12, no.~11, p.~2506, 2023.

\bibitem{Andronikos2023a}
T.~Andronikos and A.~Sirokofskich, ``A quantum detectable byzantine agreement
  protocol using only {EPR} pairs,'' {\em Applied Sciences}, vol.~13, no.~14,
  p.~8405, 2023.

\bibitem{Andronikos2023b}
T.~Andronikos and A.~Sirokofskich, ``One-to-many simultaneous secure quantum
  information transmission,'' {\em Cryptography}, vol.~7, no.~4, p.~64, 2023.

\bibitem{Karananou2024}
P.~Karananou and T.~Andronikos, ``A novel scalable quantum protocol for the
  dining cryptographers problem,'' {\em Dynamics}, vol.~4, no.~1, pp.~170--191,
  2024.

\bibitem{Andronikos2024}
T.~Andronikos and A.~Sirokofskich, ``A quantum approach to news verification
  from the perspective of a news aggregator,'' {\em Information}, vol.~15,
  no.~4, p.~207, 2024.

\bibitem{Theocharopoulou2019}
G.~Theocharopoulou, K.~Giannakis, C.~Papalitsas, S.~Fanarioti, and
  T.~Andronikos, ``Elements of game theory in a bio-inspired model of
  computation,'' in {\em 2019 10th International Conference on Information,
  Intelligence, Systems and Applications ({IISA})}, pp.~1--4, {IEEE}, jul 2019.

\bibitem{Kastampolidou2020a}
K.~Kastampolidou, M.~N. Nikiforos, and T.~Andronikos, ``A brief survey of the
  prisoners' dilemma game and its potential use in biology,'' in {\em Advances
  in Experimental Medicine and Biology}, pp.~315--322, Springer International
  Publishing, 2020.

\bibitem{Kostadimas2021}
D.~Kostadimas, K.~Kastampolidou, and T.~Andronikos, ``Correlation of biological
  and computer viruses through evolutionary game theory,'' in {\em 2021 16th
  International Workshop on Semantic and Social Media Adaptation {\&}
  Personalization ({SMAP})}, {IEEE}, 2021.

\bibitem{Kastampolidou2020}
K.~Kastampolidou and T.~Andronikos, ``A survey of evolutionary games in
  biology,'' in {\em Advances in Experimental Medicine and Biology},
  pp.~253--261, Springer International Publishing, 2020.

\bibitem{Kastampolidou2021}
K.~Kastampolidou and T.~Andronikos, ``Microbes and the games they play,'' in
  {\em {GeNeDis} 2020}, pp.~265--271, Springer International Publishing, 2021.

\bibitem{Kastampolidou2023}
K.~Kastampolidou and T.~Andronikos, ``Game theory and other unconventional
  approaches to biological systems,'' in {\em Handbook of Computational
  Neurodegeneration}, pp.~163--180, Springer International Publishing, 2023.

\bibitem{Papalitsas2021}
C.~Papalitsas, K.~Kastampolidou, and T.~Andronikos, ``Nature and
  quantum-inspired procedures {\textendash} a short literature review,'' in
  {\em {GeNeDis} 2020}, pp.~129--133, Springer International Publishing, 2021.

\bibitem{Adam2023}
S.~Adam, P.~Karastathis, D.~Kostadimas, K.~Kastampolidou, and T.~Andronikos,
  ``Protein misfolding and neurodegenerative diseases: A game theory
  perspective,'' in {\em Handbook of Computational Neurodegeneration},
  pp.~863--874, Springer International Publishing, 2023.

\bibitem{wootters1982single}
W.~K. Wootters and W.~H. Zurek, ``A single quantum cannot be cloned,'' {\em
  Nature}, vol.~299, no.~5886, pp.~802--803, 1982.

\bibitem{Nielsen2010}
M.~A. Nielsen and I.~L. Chuang, {\em Quantum computation and quantum
  information}.
\newblock Cambridge University Press, 2010.

\bibitem{Yanofsky2013a}
N.~S. Yanofsky and M.~A. Mannucci, {\em Quantum Computing for Computer
  Scientists}.
\newblock Cambridge University Press, 2013.

\bibitem{Wong2022}
T.~G. Wong, {\em Introduction to classical and quantum computing}.
\newblock Rooted Grove, 2022.

\bibitem{Cruz2019}
D.~Cruz, R.~Fournier, F.~Gremion, A.~Jeannerot, K.~Komagata, T.~Tosic,
  J.~Thiesbrummel, C.~L. Chan, N.~Macris, M.-A. Dupertuis, and
  C.~Javerzac-Galy, ``Efficient quantum algorithms for {GHZ} and w states, and
  implementation on the {IBM} quantum computer,'' {\em Advanced Quantum
  Technologies}, vol.~2, no.~5-6, p.~1900015, 2019.

\bibitem{Mermin2007}
N.~Mermin, {\em Quantum Computer Science: An Introduction}.
\newblock Cambridge University Press, 2007.

\bibitem{Deng2008}
F.-G. Deng, X.-H. Li, and H.-Y. Zhou, ``Efficient high-capacity quantum secret
  sharing with two-photon entanglement,'' {\em Physics Letters A}, vol.~372,
  no.~12, pp.~1957--1962, 2008.

\bibitem{Qiskit2024}
Qiskit, ``Qiskit open-source toolkit for useful quantum.''
  \url{https://www.ibm.com/quantum/qiskit}.
\newblock Accessed: 2024-03-02.

\bibitem{Tsai2011}
C.~W. Tsai, C.~R. Hsieh, and T.~Hwang, ``Dense coding using cluster states and
  its application on deterministic secure quantum communication,'' {\em The
  European Physical Journal D}, vol.~61, no.~3, pp.~779--783, 2011.

\bibitem{Hwang2011}
T.~Hwang, C.~C. Hwang, and C.~W. Tsai, ``Quantum key distribution protocol
  using dense coding of three-qubit w state,'' {\em The European Physical
  Journal D}, vol.~61, no.~3, pp.~785--790, 2011.

\bibitem{Huang2023}
X.~Huang, W.-F. Zhang, and S.-B. Zhang, ``Efficient multiparty quantum private
  comparison protocol based on single photons and rotation encryption,'' {\em
  Quantum Information Processing}, vol.~22, no.~7, 2023.

\bibitem{Banerjee2012}
A.~Banerjee and A.~Pathak, ``Maximally efficient protocols for direct secure
  quantum communication,'' {\em Physics Letters A}, vol.~376, no.~45,
  pp.~2944--2950, 2012.

\bibitem{Joy2017}
D.~Joy, S.~P. Surendran, and M.~Sabir, ``Efficient deterministic secure quantum
  communication protocols using multipartite entangled states,'' {\em Quantum
  Information Processing}, vol.~16, no.~6, 2017.

\bibitem{Song2019}
X.~Song, A.~Wen, and R.~Gou, ``Multiparty quantum private comparison of size
  relation based on single-particle states,'' {\em IEEE Access}, vol.~7,
  pp.~142507--142514, 2019.

\bibitem{Wolf2021}
R.~Wolf, {\em Quantum Key Distribution}.
\newblock Springer International Publishing, 2021.

\bibitem{Renner2023}
R.~Renner and R.~Wolf, ``Quantum advantage in cryptography,'' {\em {AIAA}
  Journal}, vol.~61, no.~5, pp.~1895--1910, 2023.

\bibitem{Neigovzen2008}
R.~Neigovzen, C.~Rod{\'{o}}, G.~Adesso, and A.~Sanpera, ``Multipartite
  continuous-variable solution for the byzantine agreement problem,'' {\em
  Physical Review A}, vol.~77, no.~6, p.~062307, 2008.

\bibitem{Feng2019}
Y.~Feng, R.~Shi, J.~Zhou, Q.~Liao, and Y.~Guo, ``Quantum byzantine agreement
  with tripartite entangled states,'' {\em International Journal of Theoretical
  Physics}, vol.~58, no.~5, pp.~1482--1498, 2019.

\bibitem{Wang2022a}
W.~Wang, Y.~Yu, and L.~Du, ``Quantum blockchain based on asymmetric quantum
  encryption and a stake vote consensus algorithm,'' {\em Scientific Reports},
  vol.~12, no.~1, 2022.

\bibitem{Yang2022}
Z.~Yang, T.~Salman, R.~Jain, and R.~D. Pietro, ``Decentralization using quantum
  blockchain: A theoretical analysis,'' {\em IEEE Transactions on Quantum
  Engineering}, vol.~3, pp.~1--16, 2022.

\bibitem{Qu2023}
Z.~Qu, Z.~Zhang, B.~Liu, P.~Tiwari, X.~Ning, and K.~Muhammad, ``Quantum
  detectable byzantine agreement for distributed data trust management in
  blockchain,'' {\em Information Sciences}, vol.~637, p.~118909, 2023.

\bibitem{Ikeda2023c}
K.~Ikeda and A.~Lowe, ``Quantum protocol for decision making and verifying
  truthfulness among n‐quantum parties: Solution and extension of the quantum
  coin flipping game,'' {\em IET Quantum Communication}, vol.~4, no.~4,
  pp.~218--227, 2023.

\bibitem{huttner1995quantum}
B.~Huttner, N.~Imoto, N.~Gisin, and T.~Mor, ``Quantum cryptography with
  coherent states,'' {\em Physical Review A}, vol.~51, no.~3, p.~1863, 1995.

\bibitem{lutkenhaus2000security}
N.~L{\"u}tkenhaus, ``Security against individual attacks for realistic quantum
  key distribution,'' {\em Physical Review A}, vol.~61, no.~5, p.~052304, 2000.

\bibitem{brassard2000limitations}
G.~Brassard, N.~L{\"u}tkenhaus, T.~Mor, and B.~C. Sanders, ``Limitations on
  practical quantum cryptography,'' {\em Physical review letters}, vol.~85,
  no.~6, p.~1330, 2000.

\end{thebibliography}

\end{document}